\documentstyle[10pt,aaspp4]{article}

\tighten
\def\fbf{{\bf f}}
\def\kbf{{\bf k}}
\def\xbf{{\bf x}}

\def\Abf{{\bf A}}
\def\Bbf{{\bf B}}
\def\Cbf{{\bf C}}
\def\Dbf{{\bf D}}
\def\Ebf{{\bf E}}
\def\Ibf{{\bf I}}
\def\Nbf{{\bf N}}
\def\Pbf{{\bf \Psi}}
\def\Phbf{{\bf \Phi}}
\def\Rbf{{\bf R}}
\def\Sbf{{\bf S}}

\def\Zbf{{\bf Z}}
\def\trace{{\rm{tr}}}
\def\like{{\cal{L}}}
\def\mathrm{\rm}
\def\simless{\lesssim}
\def\simgreat{\gtrsim}
\def\bs{{\bf s}}
\def\br{{\bf r}}
\def\bx{{\bf x}}
\def\avg#1{{\langle {#1} \rangle}}
\begin{document}

\slugcomment{Accepted for publication in The Astrophysical Journal}

\lefthead{Vogeley \& Szalay}
\righthead{Eigenmode Analysis}

\title{Eigenmode Analysis of Galaxy Redshift Surveys\\
I. Theory and Methods}

\author{Michael S. Vogeley\altaffilmark{1} \& Alexander S. Szalay}

\affil{Department of Physics and Astronomy\\
Johns Hopkins University\\
Baltimore, MD 21218\\
Email: vogeley@stsci.edu, szalay@pha.jhu.edu}

\altaffiltext{1}{Current address:\\
Space Telescope Science Institute\\
3700 San Martin Drive\\
Baltimore, MD 21218}

\begin{abstract}
We describe a method for estimating the power spectrum of density
fluctuations from galaxy redshift surveys that yields improvement in
both accuracy and resolution over direct Fourier analysis.  The key
feature of this analysis is expansion of the observed density field in
the unique set of statistically orthogonal spatial functions which
obtains for a given survey's geometry and selection function and the
known properties of galaxy clustering (the Karhunen-Lo\`eve
transform).  
Each of these {\it eigenmodes} of the observed density field
optimally weights the data to yield the cleanest (highest signal/noise)
possible measure of clustering power as a function of wavelength scale
for any survey.  
Using Bayesian methods, we simultaneously estimate
the mean density, power spectrum of density fluctuations, and redshift
distortion parameters that best fit the observed data.  
This method is particularly important for analysis of
surveys with small sky coverage, that are comprised of disjoint
regions (e.g., an ensemble of pencil beams or slices), or that have
large fluctuations in sampling density.
We present algorithms for practical application of this technique
to galaxy survey data.
\end{abstract}


\section{Introduction}

Recent measurements of galaxy clustering from redshift surveys and angular
catalogs, together with limits on the clustering of mass implied by
the COBE DMR experiment, yield important constraints on proposed models
for the formation of large-scale structure.
However, we lack accurate constraints on fluctuations in galaxy density
on scales that overlap with those probed by COBE, and the extant
measurements have poor resolution on scales where certain theories
predict interesting features in the power spectrum.
Several surveys, either planned or in progress, promise to yield the
desired measurements of the power spectrum of galaxy density fluctuations,
but the complex geometry and sampling of these surveys pose a strong
challenge to traditional methods of power spectrum analysis.
The ultimate measurement of the galaxy fluctuation spectrum will result
from combining all of the available data into one sample. 
This possibility begs the question,
how do we obtain the best possible estimate of the power spectrum
from a sample with arbitrarily complex geometry and with varying sampling
density? 
In this paper we describe a method for power spectrum estimation
that is optimal for any survey.
Subsequent papers in this series will describe the results of applying 
these techniques to observations.

\subsection{Why Measure the Power Spectrum?}

The power spectrum or its Fourier transform, the autocorrelation function,
measures the lowest order departures from homogeneity. Standard models
for the formation of large-scale structure provide strong motivation for
measuring the power spectrum: if structure grows via gravitational instability
from an initially Gaussian density field (as predicted by inflation), then
large-scale fluctuations at the present epoch could reveal signatures of the
initial conditions.

We choose to focus on the power spectrum rather than the correlation
function or other measures of variance because, although the power
spectrum and correlation function form a Fourier transform pair, the
former more clearly reflects the physical scales and processes that
affect structure formation on large scales.  For example, in CDM
models, the horizon size at the epoch of equality between the density
of matter and radiation is revealed by the peak of the power spectrum.
A large baryon content in the universe would cause small
``wiggles'' near this scale that might be seen in the power spectrum.
Such features would be integrated over and therefore difficult to detect in
the correlation function. 

In addition to more clearly reflecting the initial conditions,
power spectrum estimation has several statistical advantages.
Because the power spectrum of any statistical process 
is positive definite for every wavenumber,
we obtain a quick sanity check on our calculations; a negative value
tells us that, e.g., we have erred in subtracting out the shot noise
component of the power.
For likelihood analysis of proposed models, this bound usefully constrains
the available parameter space of power spectrum models.
We note that
one formerly-tauted advantage of power spectrum analysis no longer applies;
improved estimators for the correlation function 
(Landy \& Szalay 1993; Hamilton 1993) 
are afflicted by uncertainty in
the mean density only to the same degree as the standard power spectrum
estimator.

In our discussion of eigenmode analysis, we often mathematically
represent the galaxy density fluctuations with the correlation
function. Nevertheless, the quantity we seek to estimate is the power
spectrum, which we Fourier transform to compute quantities in real space,
such as the correlation function.

\subsection{Standard Methods and Current Results}

Using standard estimation techniques (as described in section 1.3 below),
the 3-D redshift-space 
power spectrum has been estimated for redshift samples of
optically-selected (CfA1 [Baumgart \& Fry 1991], SSRS1 [Park, Gott,
\& da Costa 1992], CfA2 [Vogeley et al 1992; Park et al. 1994], 
combined SSRS2+CfA2
[da Costa et al. 1994], Las Campa\~nas [Lin et al. 1995]), infrared
(IRAS 1.2Jy [Fisher et al. 1993], QDOT [Feldman, Kaiser, \& Peacock 1994,
FKP hereafter]), 
and radio galaxies (Peacock \& Nicholson 1991).
(See Vogeley 1995 for a recent review.)
These redshift-space 
power spectra all roughly agree in shape, $P(k)\propto k^n$ with
$n\approx -2$ on scales $2\pi/k< 30h^{-1}$ Mpc and $n\approx -1$ for
$30 \simless 2\pi/k \simless 120 h^{-1}$ Mpc, with weak evidence for a
turnover on a scale $2\pi/k \sim 200h^{-1}$ Mpc.
The amplitudes of these power spectra vary systematically 
with the species of galaxy in the sample.
In particular, there is
mounting evidence for luminosity bias, in the sense that bright 
optically-selected galaxies have a larger clustering amplitude than
their fainter companions, though with the same power spectrum shape
(Park et al. 1994).

These power spectrum measurements yield excellent constraints
on models with CDM-like power spectra, but are insufficient to 
differentiate among the broad classes of contending models.
The data can be well fit by a CDM power spectrum with $\Omega h \approx 0.25$
(Kofman, Gnedin, \& Bahcall 1993; Peacock \& Dodds 1994).
The power spectrum of the 
``standard'' $\Omega=1, H_0=50$kms$^{-1}$ model is excluded due to an excess
of small vs. large-scale power.
Due to the strong influence of peculiar velocities in this model, 
the shape of the redshift-space power spectrum is roughly correct, but
the amplitude is too high
(when normalized to COBE, this model
requires {\it anti-biasing} of all but the very brightest galaxies
[Stompor, G\'orski, \& Banday 1995]).
However, several alternative models predict power spectra with nearly the same
shape and the correct normalization,
among which the current data do not strongly discriminate.
The list of candidates includes (but is not limited to) 
CDM models with non-zero cosmological constant, open universe
CDM, mixed (cold plus hot) dark matter models (e.g., Primack et al. 1995),
and warm plus hot
dark matter (e.g., Malaney, Starkmane, \& Widrow 1995).

To further constrain cosmological models, we must (1) close the gap
between the scales probed by galaxy surveys and COBE, (2) measure the
detailed shape of the galaxy power spectrum, (3) determine the
dependence of clustering on galaxy species, and (4) quantify the anisotropy
of clustering in redshift space caused by peculiar motions of galaxies.
Comparison of power spectra for currently competing models 
(e.g., Figure 1 of Strauss et al. 1995)
shows 
that the shapes of these spectra differ most greatly on scales near
and beyond the peak of the spectrum.
There are hints of features in the power spectrum (e.g., the
feature at $\lambda\sim 30h^{-1}$ Mpc in the CfA2 and SSRS2 power
spectra [da Costa et al. 1994] and the peaks at $\lambda \sim 30h^{-1}$ and
$128h^{-1}$ Mpc seen by Broadhurst et al. 1990). The data are
consistent with a turnover on large scales to a $n=1$ spectrum that
would be consistent with COBE.
However,
more accurate probes of scales $\sim 100h^{-1}$ Mpc and greater are necessary
to test for features in the power spectrum on scales where physical processes
near the time of matter-radiation equality would leave their imprint and,
ultimately, to compare galaxy clustering and the amplitude of mass clustering
implied by CMB anisotropy measurements.
The latter, along with knowledge of the dependence of clustering on
galaxy selection, will elucidate the relationship between clustering of 
mass and light in the universe.
Furthermore, measurement of the anisotropy of clustering in redshift space
on the largest (and, therefore, presumably linear-growth) scales can yield
a direct measurement of the mean cosmic density
(Kaiser 1987; Hamilton 1992; Cole, Fisher, \& Weinberg 1993).

\subsection{Improved Data Demand Improved Analysis:
Problems with Standard Methods}

To increase the largest scales that we probe and the resolution of
these measurements, we must survey a larger volume of the universe.
Several ongoing and planned surveys promise to yield better
constraints on the fluctuation spectrum, but the geometries of these
samples pose a challenge to standard methods of power spectrum
estimation.  Deeper surveys of this type that are completed, or are
soon to be, include pencil beam surveys
(Broadhurst et al. 1990, 1995)
and several deep slice surveys: the Las Campa\~nas
(Shectman et al. 1995), Century (Geller et al. 1995), and ESP
(Vettolani et al. 1995) surveys.  Within the next two years we also
expect results from the AAT 2df survey and the Sloan Digital
Sky Survey (SDSS hereafter).  Most of the sensitivity of the AAT
survey to large-scale fluctuations
will result from an ensemble of 100 randomly spaced pencil
beams of 400 galaxies each.  Over its five year duration, the SDSS
will obtain redshifts for $10^6$ galaxies over a contiguous area of
$\pi$ steradians in the North Galactic Cap, and therefore have a
rather simple geometry, but earlier partial data (e.g., $2\times 10^5$
galaxies over some set of narrow stripes on the sky in the first year)
and the survey in the South (three $2\fdg 5\times 100\arcdeg$ stripes) will be
more complex.

Standard methods for estimating the power spectrum all follow the
same basic scheme: We directly
sum the planewave contributions from each galaxy,
\begin{equation}
\tilde{\delta}(\kbf)={1\over \sum_j w(\xbf_j)}
	\sum_j w(\xbf_j) e^{i\kbf \cdot \xbf_j} - \tilde{W}(\kbf),
\end{equation}
where $w(\xbf_j)$ is the weight given to the $j^{th}$ galaxy and we
subtract $\tilde{W}(\kbf)$, 
which is the contribution to each mode from the finite survey window 
($W(\xbf)=1$ inside the survey and 0 elsewhere)
and the selection function $\bar{n}(\xbf)$,
\begin{equation}
\tilde{W}(\kbf)= { \int d^3x\: w(\xbf)\bar{n}(\xbf)W(\xbf) e^{i\kbf\cdot\xbf}
\over \int d^3x\: w(\xbf)\bar{n}(\xbf)W(\xbf) }.
\end{equation}
Next we compute the square of the modulus of each Fourier coefficient
and subtract the power due to shot noise, 
\begin{equation}
\hat{P}(\kbf)= {|\tilde{\delta}(\kbf)|^2 - 
\sum_j w^2(\xbf_j)\,/\,\left[\sum_j w(\xbf_j)\right]^2
\over {1\over (2\pi)^3} \int d^3k' \, |\tilde{W}(\kbf')|^2},
\end{equation}
and average these estimates over a shell in k-space to yield an estimate
of $P(k)$.
The denominator of equation (3) enforces the convention that $P(k)$ has the
units of volume.
Methods vary in the details (compare Park et al. 1994, Fisher et al. 1993,
and FKP), including the weights applied to each galaxy, how the
window function of the survey is computed, corrections (or lack thereof)
for the damping of large-scale power (analogous to the integral constraint
on the correlation function -- see below), 
and attempts to deconvolve the true power
from the window function of the survey.

The standard methods of power spectrum analysis have several weaknesses
that become even more serious when applied to surveys with complex
geometry and sampling.
A critical problem is
that the basis functions of the Fourier expansion (plane waves) are not
orthonormal over a finite non-periodic volume. 
Following equations (1)-(3)
the power measured at any wavenumber is a convolution of the true
power with the
window function of the survey,
\begin{equation}
\hat{P}(\kbf)=\int d^3k'\, P(\kbf') |W(\kbf-\kbf')|^2 .
\end{equation}
The estimates of power at different wavenumber have a covariance
that depends on the shape of the survey volume.
If the survey is oddly-shaped, then estimates $\hat{P}(\kbf)$ with the
same $k=|\kbf|$ but different direction $\hat{\kbf}$ sample different
ranges of wavenumber because $W(\kbf)$ is anisotropic.
Averaging over a shell in $k$-space combines power estimates with
varying bandpass and, therefore, different signal-to-noise ratios.

Further complications arise when we consider how to optimally 
weight the galaxies (as in eq. [1]) in different regions of a survey. 
The signal-to-noise for detection of
clustering depends on the sampling rate of galaxies in the survey, which
may vary due to survey strategy (e.g., the Las Campa\~nas survey,
which observes the same number of galaxies in each plug plate field),
extinction (for a survey which includes galaxies to a fixed apparent
magnitude if uncorrected for extinction), combining different surveys
into a single sample, or simply because the selection function varies with
distance (in the case where we analyze apparent-magnitude
limited samples). 
FKP derive a weighting scheme that yields minimal
variance in the standard power spectrum estimator,
\begin{equation}
w(\xbf)={1\over 1+\bar{n}(\xbf)P(k)}.
\end{equation}
This weighting scheme is correct in the case where (1) the density
fluctuations are Gaussian, 
(2) the only source of uncertainty is shot noise, and 
(3) the window function is
close to spherically symmetric (the latter is, in fact, the case for
the QDOT sample that they examine).

In general, the set of weights applied to the galaxies should
vary with the wavenumber $\kbf$ being probed.
In practice, most authors use a single set of weights, following
the argument that it is tedious to vary them with $\kbf$
(because it requires
an {\it a priori} estimate of $P(\kbf)$ at each wavenumber)
and that the results do not depend sensitively on varying $P(\kbf)$
in the weights.
If the sampling density strongly varies, then this approximation
is quite poor.
In the case of the FKP analysis of the QDOT survey,
it is clear that the estimated $P(k)$ does vary with the
weighting scheme applied. Because they examine a flux-limited sample,
the weighting scheme determines the effective depth of the
volume used to probe the fluctuations.
If the sampling density of galaxies varies with position on the
sky, as in the Las Campa\~nas survey, then the variation with wavenumber
of the weight per galaxy yields a different pattern on the sky
for each mode.
Ignorance of this variation with wavenumber of the weighting scheme
yields estimates of power with unnecessarily
poor signal to noise.

Uncertainty in the mean density limits our ability to detect fluctuations
on very large scales.
In equation (1), when we subtract the
contribution of the window function to each Fourier mode,
we attempt to subtract the spike at $k=0$ in the true
power, which is due to the non-vanishing mean density of galaxies.
Because this spike is convolved with the window function of the
survey and because we typically estimate the mean density from the sample
itself (which forces $\langle \delta(\xbf)\rangle=0$ within the
survey), we erroneously subtract the product of the window function
with the component of clustering signal on the scale of the survey,
$|W(\kbf)|^2\langle |\delta_0|^2\rangle$.
In other words, we underestimate clustering on scales comparable to
and larger than the survey because we cannot (rather, do not
attempt to) ascertain if our chosen
volume is under or overdense relative to the rest of the universe.
It is possible to correct for this damping of power on large scales
(Peacock \& Nicholson 1991; Park et al. 1994),
but only if we know the true power spectrum.
Odd geometry only complicates this correction:
if the geometry and, therefore, the Fourier window is anisotropic, 
then this power damping will be different for each mode.
If the survey volume has elements that are narrow in any direction, then
the mean density problem will extend down to relatively smaller scales.
A better method would be to simultaneously estimate both the mean
density and the power spectrum.

Following standard methods, model testing is made difficult by the
non-orthonormality of the Fourier modes, ambiguity about the optimal
number of modes to include in the analysis, and necessary assumptions
about the probability distribution of the measured power per mode.  In
principle, we could test models by computing the full covariance
matrix of the power spectrum estimates for each model in consideration
and compare their likelihoods, as approximated in FKP.
This procedure requires that we repeatedly invert large,
highly nondiagonal matrices. 
The size of the matrices could be reduced if we choose a limited set
of modes, but this method does not specify the optimal set of power
estimates; nearby modes have large covariance, but
we lose statistical power and resolution if we sample too few.
Finally, this method
requires that we know
the covariance matrix of the {\it power} per mode (which depends on 
fourth-order moments of the galaxy density) and the probability
distribution of these fourth-order fluctuations for every model under
consideration.
To test the likelihood that the observations arise from
a model with a particular power spectrum,
we only require prediction of the covariance matrix
of the expansion coefficients themselves,
and knowledge of the probability distribution of second-order fluctuations
in the density.
A good choice of basis functions eases calculation of this matrix
and increases the statistical power of the likelihood function.

\subsection{Optimal Probes of Spatial Clustering: the Karhunen-Lo\`eve
Transform}

Rather than make small modifications to the standard method for
power spectrum estimation, in this paper
we begin anew and derive the complete set
of spatial functions that optimally weight the observed data in order
to estimate second-order clustering properties of the galaxy distribution,
and describe how this expansion naturally leads to straightforward
likelihood analysis of proposed models. 

The problem of deriving a set of orthonormal functions that are optimal
for representing data with known statistical properties has been studied
in detail by investigators in the field of signal processing.
For any second-order (mean-square integrable) statistical
process, a unique set of orthonormal functions can be found such that
the expansion coefficients in this basis are statistically orthogonal
(see section 2.1 below for discussion of the differences between
statistically {\it orthogonal, uncorrelated}, and {\it independent}).
Expansion of an observed data set in this unique set of functions, or
{\it eigenmodes}, is known as the
Karhunen-Lo\`eve transform (see, e.g., Therrien 1992 or Poor 1994 for 
discussion of the discrete and continuous transforms, respectively).
In its discrete form, this transform proves useful for image compression,
filtering and, as we shall discuss, for testing models that predict the
second-order clustering properties of the observations.
In a nutshell (see section 2 for a detailed description), the Karhunen-Lo\`eve
transform uses our {\it a priori} knowledge of noise, clustering, and
geometry to derive a unique orthonormal basis set for representing
the fluctuations in each survey.
Because these eigenmodes form a complete basis, are statistically
orthogonal, and maximize the signal-to-noise per mode, representation
in this basis is optimal for testing the likelihood of proposed
clustering models.
This transform simultaneously addresses the problems of forming an orthonormal
basis and deriving optimal weights for the data.
In Vogeley (1995) we introduce the Karhunen-Lo\`eve transform as a tool
for probing density fluctuations in the galaxy distribution. In this paper
we describe in detail how we use this method to simultaneously estimate
the mean density, power spectrum, and redshift-space distortions
from galaxy redshift surveys.

In section 2 we derive the Karhunen-Lo\`eve transform and several of its
important properties. 
In section 3 we show how to derive the eigenmodes for a 
galaxy redshift survey and investigate how these modes form an optimal
set of filters for power spectrum estimation.
Section 4 provides a brief introduction to model testing in the
Bayesian paradigm and explains how we estimate the confidence regions
for proposed models.
In section 5 we summarize our estimation method,
compare with other transform methods,
and describe plans and predictions 
for application of our eigenmode method 
to existing and forthcoming survey data. 
In Appendix A we show that the Karhunen-Lo\`eve transform is the optimal
basis set for testing clustering models.
In Appendix B we derive an approximate method for computing the integral
average of the correlation function between two cells.

\section{The Karhunen-Lo\`eve Transform}

\subsection{Definition and Properties}

Because the Fourier modes are not orthonormal over the finite volume of the
survey, this transform is not ideal for representing the observed
distribution of galaxies.  
One can construct an infinite number of alternative basis functions
that are orthonormal over the survey geometry, but the
optimal choice of basis depends on the
statistical question that we ask of the data.  We want to test models
for the galaxy distribution that predict the expectation value and
second moments of the observed density. Therefore we should expand the
observed density field in a set of orthonormal functions that weight
the data to yield optimal signal to noise for the second moment of
the density and for which the
expansion coefficients are statistically orthogonal (these two
conditions turn out to yield the identical set of functions).  These
requirements on the basis set of the expansion yield a unique set of
spatial filters that are optimal for, among other purposes, power
spectrum estimation.  In this section we describe how to derive this
set of eigenmodes for any scalar function $f(\xbf)$ over the survey volume;
later we must decide which
scalar field that should be, e.g., the observed number counts of
galaxies or the density contrast of the number counts.

To allow us to present the mathematics in compact form using
matrix algebra, and to expedite the implementation of this method on
a computer, suppose that 
we divide the survey volume into $M$ cells, each with 
volume $V_i$.
The $i^{th}$ cell is centered at $\xbf_i$ and 
we measure $f(\xbf_i)$ in each cell.

Using matrix notation, a scalar function $f(\xbf_i)$ is a vector
$\fbf$, which we can expand in a set of $M$
orthonormal basis vectors $\{\Pbf_n(\xbf_i); n=1,M\}$
with vector of coefficients $\Bbf$,
\begin{equation}
\fbf = \Pbf\Bbf,
\end{equation}
where the vectors $\Pbf_n$ form the columns of the matrix $\Pbf$.
The coefficients of the expansion are defined by the transform
\begin{equation}
\Bbf = \Pbf^{-1} \fbf.
\end{equation}
The orthonormality condition is
\begin{equation}
%%	\cdot denotes the inner product of \fbf
\Pbf_i^*\cdot \Pbf_j = \delta_{ij}.
\end{equation}
With this condition on the basis set, $\Pbf$ is a unitary matrix and
this expansion is
equivalent to a rotation
of $\fbf$ into the space spanned by the set of basis vectors $\{\Pbf_n\}$.
The inverse of a unitary matrix is its adjoint, thus $\Pbf^{-1}=\Pbf^{\dag}$.

When we test the likelihood of the observed expansion coefficients,
it will prove useful if the correlation matrix
is as diagonal as possible, thus we
impose the further condition that the expansion coefficients in this basis
be statistically orthogonal,
\begin{eqnarray}
\langle B_i B_j^* \rangle
& = &\langle (\Pbf_i^{\dag} \fbf) 
(\Pbf_j^{\dag} \fbf)^{\dag} \rangle \nonumber \\
%% 	note that \circ here denotes the outer product of \fbf
& = & \Pbf_i^{\dag} \langle \fbf \circ \fbf \rangle \Pbf_j \nonumber \\
& = & \langle B_i^2 \rangle \delta_{ij},
\end{eqnarray}
which
implies that the basis functions that we seek solve the eigenvalue problem
\begin{equation}
\Rbf \Pbf_j = \lambda_j \Pbf_j,
\end{equation}
where the correlation matrix
of the function $f(\xbf_i)$ has elements
$R_{ij}=\langle f(\xbf_i) f(\xbf_j) \rangle$,
the $\Pbf_j$ are the eigenvectors of this
correlation matrix, and the eigenvalues are
$\lambda_j=\langle B_j^2\rangle$.  
Expansion in the set of eigenvectors of the
correlation matrix is the discrete form of the
Karhunen-Lo\`eve transform (K-L hereafter).

It is important to clarify the differences between statistically
{\it orthogonal}, {\it uncorrelated}, and {\it independent}.
Statistical orthogonality is the condition stated in equation (9).
If $f(\xbf)$ has zero mean, then this same condition implies
that the coefficients are also uncorrelated.
For the coefficients to be statistically independent, 
we further require $f(\xbf)$ (and
therefore $B_n$) to be a Gaussian random process.
Statistical orthogonality alone 
does not require $f(\xbf)$ to be a Gaussian random process.

Here and throughout this paper, the operators $\langle \rangle$ denote
the expectation value, or ensemble average, of a quantity. Under the
assumption that galaxy clustering is an ergodic process, the ensemble
average is equivalent to a spatial average.  In other words, $\langle
n(\xbf)\rangle$ is not only the expectation value of the density in
our particular survey, but also the expectation value of the density
at that position within an identically conducted survey in another
patch of the universe.  This equivalence is less trivial than it first
appears: we are concerned not merely with uncertainty in the measured
density at a particular point in space (caused by, e.g., Poisson
fluctuations in number density and measurement errors), but also
with genuine correlated fluctuations about the cosmic mean.

Note that
the number of eigenvectors corresponds to the number of pixels with
which we divide the survey.  If the cell size is comparable to, or
smaller than, the average intergalaxy spacing, then using a finer mesh
does not change or increase the number of eigenmodes that sample
large-scale fluctuations; we merely add very low signal to noise modes
that are sensitive to small-scale pixel-to-pixel fluctuations and
are dominated by shot noise.  In section 3.1 we discuss constraints on
pixellating a galaxy survey for this analysis.

The K-L transform is unique; the eigenvectors of the
correlation matrix form the only orthonormal basis set for which the transform
coefficients $B_n$ are statistically orthogonal.  
To demonstrate this uniqueness property (Therrien 1992),
consider an arbitrary
set of orthonormal vectors $\{\Phbf_n\}$.
The condition of statistical orthogonality is as above (eq. [9]),
\begin{equation}
\langle A_i A^*_j \rangle = 
\Phbf_i^{\dag} \Rbf\Phbf_j = \langle A_i^2\rangle \delta_{ij},
\end{equation}
where the $A_i$ are the expansion coefficients in the new basis.
This condition may be rewritten as
\begin{equation}
\Phbf^{\dag}_i {\bf w}_j = \langle A_i^2 \rangle\delta_{ij},
\end{equation}
where ${\bf w}_j=\Rbf \Phbf_j$.
Each of the ${\bf w}_j$ must be orthogonal to every $\Phbf_i$ for
$i\neq j$ and the $\Phbf_i$ are orthonormal, thus ${\bf w}_j$
is simply some constant times $\Phbf_j$, i.e.,
\begin{equation}
{\bf w}_j=\Rbf \Phbf_j = \lambda_j \Phbf_j,
\end{equation}
so that $\Phbf_j$ is an eigenvector of $\Rbf$, with eigenvalue
$\lambda_j = \langle A_j^2\rangle$.
Because this is true for all $j$, the eigenvectors of the correlation
matrix are the only set of $\Phbf_j$ that are statistically
orthogonal.

Another unique property of the K-L transform
is that it yields the most efficient representation of the data if we
truncate the expansion to include fewer than $M$ modes.
To demonstrate this property,
suppose that we expand the scalar field $f(\xbf)$ in the orthonormal
basis 
$\{\Pbf_n\}$, but that we truncate the expansion at $N < M$ basis functions,
\begin{equation}
\hat{\fbf}=\sum_{i=1}^{N<M} B_i \Pbf_i^{\dag}
\end{equation}
and define the error vector $\epsilon=\fbf - \hat{\fbf}$.
The total power lost in the truncation is 
\begin{equation}
\epsilon^2= \langle \epsilon \epsilon^{\dag} \rangle
= \left \langle \left (\sum_{i=N+1}^M B_i \Pbf_i^{\dag}\right ) 
\left (\sum_{j=N+1}^M B_j^* \Pbf_j \right)\right \rangle 
= \sum_{i=N+1}^M \langle B_i^2\rangle.
\end{equation}
To minimize the lost power, we must therefore minimize
\begin{equation}
\epsilon^2=\sum_{i=N+1}^M \Pbf_i^{\dag} \Rbf \Pbf_i
\end{equation}
subject to the orthonormality constraints.
Using the method of Lagrange multipliers,
we solve this problem by minimizing the Lagrangian
\begin{equation}
\like= \sum_{i=N+1}^M \Pbf_i^{\dag} \Rbf \Pbf_i
+ \lambda_i(1-\Pbf_i^{\dag} \Pbf_i).
\end{equation}
Note that $\like=\epsilon^2$ when the orthonormality condition
of the $\Pbf_i$ is satisfied, and is minimized when
\begin{equation}
{\partial \like \over \partial\Pbf_i} = \Rbf \Pbf_i - \lambda_i \Pbf_i = 0,
\end{equation}
where $\partial\like/\partial\Pbf_i$ 
is the gradient with respect to changes in the basis
vectors $\Pbf_i$.
Again, we find that the $\Pbf_i$ must be eigenvectors of the correlation
matrix, with eigenvalues $\lambda_i=\langle B_i^2\rangle$.
The optimal basis vectors for the truncated expansion are therefore
the $N$ vectors $\Pbf_i$ with the largest eigenvalues $\lambda_i$.
This efficiency property obtains for any number of modes, and therefore
for a single mode, thus
the first eigenmode yields an optimal estimator for the mean value
of the observed field.
If we form the eigenmodes from the covariance matrix (we first
subtract an estimate of the mean field at each point) rather than the
correlation matrix then, of course, we lose this estimator.

\subsection{Signal to Noise Properties of the Eigenmodes}

If the noise in $f(\xbf_i)$ 
is white (e.g., shot noise) and we divide the survey volume so that each
cell has the same noise, then the efficiency property of the
K-L transform implies that this transform also
yields the maximum possible signal to noise per mode.
As long as the signal and noise in $f(\xbf)$ are uncorrelated,
the correlation matrix may be written as the sum of terms
$\Rbf=\Sbf+\Nbf$,
which depend on the expected
signal and noise, respectively.
For constant noise per cell, the noise correlation matrix is
$\Nbf=\sigma^2\Ibf$, the product of a scalar with the identity matrix.
The K-L transform always diagonalizes $\Rbf$, so in this case it also
diagonalizes the signal correlation matrix $\Sbf$,
so that the signal and noise remain uncorrelated in the new basis.
The noise power per mode, $\Pbf_n^{\dag} \Nbf \Pbf_n = \sigma^2$, is constant,
thus sorting the eigenvalues $\lambda_n$ is equivalent to sorting by
signal-to-noise ratio
(throughout this paper we assume that the eigenvectors have been sorted
in order of decreasing eigenvalue).
This property of the K-L transform leads to Bond's (1994) description
of these functions as ``signal-to-noise eigenmodes.''
An important consequence of representing the signal with the smallest
possible number of statistically orthogonal modes
is that the likelihood function of the coefficients discriminates as
strongly as possible between different clustering models (see Appendix A).

The K-L transform retains its signal-to-noise optimization 
property for {\it arbitrary} pixellation if we first
apply a whitening transformation to the binned measurements $f(\xbf_i)$.
Because the pixellation may be driven by requirements other than optimizing
the signal to noise,
it is important to have a prescription for deriving the
eigenmodes that does not depend on a particular division of space into cells.
If the noise per cell varies, the noise matrix $\Nbf$ is not proportional
to $\Ibf$
and a transformation that diagonalizes $\Rbf$ leaves the signal and
noise correlated in the new basis.
To ensure that this mixing does not occur, we can either make a clever choice
of pixellation (as above) or weight the cells to account for their varying
noise properties.
Before we find the eigenvectors, we prewhiten (diagonalize
the noise component of) the correlation matrix to form 
\begin{eqnarray}
\Rbf' & = & \Nbf^{-1/2} \Rbf \Nbf^{-1/2} \nonumber \\
      & = & \Nbf^{-1/2} \Sbf \Nbf^{-1/2} + \Ibf,
\end{eqnarray}
where the elements of the whitening transform
$\Nbf^{-1/2}$ are the square roots of the elements of $\Nbf^{-1}$
and $\Ibf$ is the identity matrix.
The complete K-L transform of ${\bf f}$ is then
\begin{equation}
\Bbf=\Pbf^{\dag}\Nbf^{-1/2}\fbf,
\end{equation}
where $\Pbf_n$ are the eigenvectors of $\Rbf'$.
The whitening transformation
rescales the data space to account for the differing noise per cell.
Expansion in the eigenvectors of $\Rbf'$ is a
rotation within the data space (a unitary transformation) that  
diagonalizes the signal.

The whitening transform also gives us a procedure for
generalization to more complex noise processes, in which the
noise correlation matrix is not diagonal.
In this case, the whitening transform not only rescales the data space,
but also rotates within this space to diagonalize the noise matrix.
The vectors $\Nbf^{-1/2}\Pbf_n$ are the solutions $\Phbf$ to
the generalized eigenvalue problem 
$\Sbf\Phbf=\lambda \Nbf \Phbf$
and may be found either by direct solution of this equation or via
the two step process above. 
In Appendix A, we show that the transform $\Nbf^{-1/2}\Pbf$ is the 
optimal transform for testing clustering models for any data set.

The eigenvectors that we derive on a mesh of pixels
are merely approximations to the true continuous eigenmodes,
which are continuous functions of position and are infinite in number.
For our purposes,
this approximation is sufficient when the
scale of the cells is considerably smaller than the scale of the
fluctuations that we wish to probe.

\section{Eigenmodes of a Galaxy Redshift Survey}

\subsection{Pixellation of the Survey Volume}

To implement this estimation method on a computer, we must divide the
survey volume into a finite number of cells. 
We specify the shape and volume of the cells, as well as
how these vary with position within the survey volume.  
We try to make the cells as spherically symmetric as possible, to
improve the accuracy of our approximate methods for
computing $\xi_{ij}$ (see Appendix B).

If the pixels are too large, then we lose resolution because the binning
smooths the galaxy distribution.  
Available computing resources and one's patience set a practical 
upper limit to the number of cells.
We suggest two scales to consider as useful lower
bounds on the pixel size.
One such scale is the galaxy correlation
length, $r_0\sim 5h^{-1}$Mpc,
because we are interested in studying fluctuations on
much larger scales, which have yet to be accurately probed.
The other scale of interest is
the average intergalaxy spacing
$\bar{n}^{-1/3}$.
Although the best possible results obtain for nearly infinitesimal cells, 
there are diminishing returns when the number of cells
exceeds the number of galaxies, 
because truly infinitesimal pixels admit
eigenvectors with very low signal to noise
at the cost of constructing and diagonalizing a larger correlation matrix. 

At large distance, where the selection function drops rapidly, the
appropriate cell volume quickly blows up -- this is the point
at which we should set the outer boundary for a magnitude-limited sample.
A volume-limited sample has a fixed outer boundary -- the maximum 
distance to which
our sample is complete to the chosen absolute magnitude limit
(although uncertainty in the apparent magnitudes makes this boundary
``fuzzy'').

\subsection{The Correlation Matrix}

We observe a set of galaxy counts $d_i$ 
in cells $V_i$ of a galaxy redshift survey, which form an observed
data vector $\Dbf$.
To optimally represent the fluctuations in galaxy density, we 
expand these observation in the eigenmodes of $\Dbf - \langle \Dbf \rangle$.
Rather than subtract the mean density before we apply the K-L transform,
we expand the non-zero mean $\Dbf$ in these eigenmodes.
When we test clustering models, we subtract the expectation value of
the K-L coefficients predicted by each model.
Thus reduced to a zero-mean process, all of the above results apply
for the K-L expansion of the density fluctuations.

The correlation matrix of galaxy density fluctuations in cells has elements
(see Peebles 1980 for a derivation of moments of counts-in-cells)
\begin{eqnarray}
R_{ij} & = & \left \langle (d_i-\langle d_i\rangle)
(d_j -\langle d_j\rangle) \right \rangle 
\nonumber \\
       & = & n_i n_j \xi_{ij} + \delta_{ij} n_i + \epsilon_{ij}
\end{eqnarray}
where $n_i \equiv \langle d_i\rangle$, $\delta_{ij}=0$ for $i\neq j$,
$\epsilon_{ij}$ is the correlation matrix for other sources of noise,
and
\begin{equation}
\xi_{ij}
\equiv {1\over V_iV_j}\int d^3x_i \int d^3x_j \,
\xi({\bf x_i}, {\bf x_j}).
\end{equation}
The three terms in equation (21) are the contributions from
clustering
of galaxies, shot noise, and extra variance due to, e.g., magnitude errors
or uncertainty in the luminosity function.
The correlation function includes the redshift space distortions,
$\xi(\xbf_i, \xbf_j)=\xi(r_p, \pi, R)$, 
where $r_p$ is the projected separation,
$\pi$ the line of sight separation, and R the distance of the pair
of cells from the observer.

The expected counts are
\begin{equation}
n_i = \int_{V_i} d^3x\, \bar{n}(\xbf).
\end{equation}
We require knowledge of the geometry of the cells and the selection
function for galaxies within the survey volume $\bar{n}(\xbf)$.
Number density or luminosity evolution of the galaxy population
may be included in the function $\bar{n}(\xbf)$.

To compute the correlation function in redshift space, we first compute
the real-space correlation function by Fourier transform of the real-space
power spectrum, then apply a distortion to form the redshift-space
correlation function at each position.
By explicitly including the redshift distortions as part of the clustering
model, we can simultaneously estimate both the real-space power spectrum
and the strength of these distortions. 
We could also include a function multiplying $\xi$ that describes
modulation of the clustering amplitude by, e.g., luminosity dependence,
clustering evolution, etc.

To compute the average value of the correlation function between
cells, we adopt either of two approximations,
depending on the distance between two cells.
For distant cells, we use an approximation based on expanding the correlation
function in a Taylor series and using the inertial moments of the cells.
When the cells are close enough that this approximation is no longer valid,
we compute the correlation term by Monte Carlo integration of the correlation
function between the two volumes.
We exploit symmetries in the survey geometry to compute
the correlation term for the minimal number of unique relative positions
of the cells.
Appendix B describes the Taylor series approximation for computing
$\xi_{ij}$ in the case where we ignore the redshift-space anisotropy.
However, the full power of the eigenmode method depends on accurate modelling
of this distortion. 
We will describe a method for modelling
the effects of this anisotropy and an approximation for computing
$\xi_{ij}$ (without using
the ``distant-observer'' approximation) in a future paper in this series.

For the case where shot noise is the only source of noise, the noise
correlation matrix has the simple form $N_{ij}=n_i\delta_{ij}$.
To be complete, we should also take into account other sources of uncertainty
in the observed counts including,
for example, magnitude errors and uncertainty in corrections for
Galactic extinction.
Our procedure for constructing the eigenmodes optimally weights the
data for varying signal to noise.
In the case of correlated noise, the 
noise correlation matrix is not diagonal, 
in which case the whitening transformation
is particularly useful for isolating the uncorrelated set of observed
counts.

Note that all these calculations implicitly assume a cosmological model
and a local flow model to convert redshifts to comoving coordinates.
To be consistent, we must
use the same coordinate transformation at each step (computing
the distance to each galaxy, forming the eigenmodes, computing the K-L 
coefficients, forming the correlation matrix of coefficients for proposed
models).
Because the calculation of comoving coordinate distances from redshifts
depends on the assumption that the observer is at rest, we should
correct the observed redshifts for our own peculiar motion.
Failure to correct to the proper reference frame can yield an erroneous
anisotropy in the galaxy distribution, the so-called ``rocket effect''
(Kaiser 1987).
For a very shallow survey, we might want to work in the Local Group reference
frame, and therefore only correct for our motion with respect to Virgo.
For much deeper surveys, it makes sense to work in the frame in which the
CMB is isotropic.

\subsection{Finding the Eigenmodes}

Before we find the eigenvectors of the correlation matrix, we
apply a whitening transformation, as described above in section 2.2
and in Appendix A.
If shot noise is the only source of noise, then the noise correlation
matrix has elements $N_{ij}=\delta_{ij}n_i$, with inverse
$N^{-1}_{ij}=\delta_{ij}/n_i$.
The whitened correlation matrix 
$\Rbf' = \Nbf^{-1/2}\Rbf\Nbf^{-1/2}$
has elements 
\begin{eqnarray}
R'_{ij} & = & n_i^{-1/2}[n_i n_j \xi_{ij} + \delta_{ij}n_i]n_j^{-1/2}
\nonumber \\
& = & n_i^{1/2}n_j^{1/2} \xi_{ij} +\delta_{ij}
\end{eqnarray}
Note that we obtain an identical correlation matrix $\Rbf'$
if we first weight
each galaxy by the inverse of the selection function (or any other
{\it a priori} set of weights)
because this weighting
is removed by the whitening transformation. 

We find the eigenvectors $\Pbf_n$ of the whitened correlation matrix,
which yield the
K-L transform of the data, with expansion coefficients
$B_n=\Pbf_n^{\dag} \Nbf^{-1/2} \Dbf$.
The eigenvalues $\lambda_n$ are
the expectation value of the square of the coefficients
$\lambda_n=\langle B_n^2\rangle$.
The elements $\Psi_n(\xbf_i)$ of the eigenvectors multiplied by the
whitening transformation $\Nbf^{-1/2}$,
specify the weight given to the $i^{th}$ 
cell for the $n^{th}$ eigenmode. 
If shot noise is the only source of uncertainty, then
$\Nbf^{-1/2}\Pbf_n(\xbf_i)=\Pbf(\xbf_i)/(\bar{n}(\xbf_i)V_i)^{1/2}$.
From these weights per cell we can form the continuous function of
position,
\begin{equation}
F_n(\xbf)=\Psi_n(\xbf_i)/V_i^{1/2},
\end{equation}
for $\xbf\in V_i$,
that obeys the orthonormality condition,
\begin{eqnarray}
\int d^3x \, F_m(\xbf) F_n(\xbf)& =& \sum_i V_i (\Psi_m(\xbf_i)V_i^{-1/2})
(\Psi_n(\xbf_i)V_i^{-1/2})\\
&=&\sum_i \Psi_m(\xbf_i) \Psi_n(\xbf_i)
=\Psi_m\cdot\Psi_n=\delta_{mn}.\nonumber
\end{eqnarray}

We use the first slice of the CfA2
redshift survey (de Lapparent, Geller, \& Huchra 1986) to illustrate
our methods.
This survey slice covers the region
$29\fdg 5\leq\delta\leq 32\fdg 5$, $8^h\leq \alpha\leq 17^h$,
and we restrict our example to $10h^{-1}{\mathrm{Mpc}}\leq r\leq
120h^{-1}{\mathrm{Mpc}}$.
We divide the slice into a single layer of $M=1225$
pixels (slightly more pixels than galaxies) in spherical coordinates and
compute the correlation matrix of expected galaxy counts (not the
observed counts) in this apparent-magnitude limited sample using the
selection function of this survey and assuming 
the power spectrum measured from the full CfA2 survey (Park et al. 1994).
We then find the eigenmodes by whitening this correlation matrix and
finding its eigenvectors, as described above.
For the test case of the CfA2 slice that we describe, we use
the Jacobi transform method to compute the eigenvectors and eigenvalues, 
as described in Press et al. (1986).
A more reliable (slower, but safer) method is Singular Value Decomposition.
Standard linear algebra packages (e.g., LAPACK -- Anderson et al. 1995) 
include a number
of routines for finding the eigenvectors and eigenvalues of a real,
symmetric matrix, as is the case for the correlation matrices.

Figure 1 shows the most significant (largest eigenvalue) 12 eigenmodes.
We plot the discrete approximation to the continuous eigenfunctions,
$F_n(\xbf)$, as in equation (25).
These modes resemble dipole, quadrapole, etc., 
moments of the galaxy distribution in the slice. 
Because these are eigenmodes of the magnitude-limited galaxy counts, 
the amplitude of each mode varies with depth so that
it is most sensitive to fluctuations near the peak of the expected
redshift distribution ($N(r)\propto r^2\bar{n}(r)$ peaks near 55$h^{-1}$Mpc
and approaches zero
sensitivity beyond $r\sim 100h^{-1}$Mpc.

Figure 2a shows the familiar distribution of galaxies in the CfA2
slice, but we bin the galaxies into pixels rather than show the galaxy
positions themselves. 
Figure 2b illustrates the optimal representation property of the K-L transform;
here we reconstruct the galaxy counts using only the first 500 eigenmodes of
the survey, yet all the salient features are reproduced. 
The error image (Fig. 2c) shows the difference between the true and
truncated distribution, which is formed by the remaining
725 eigenmodes, all of which have signal-to-noise of less than unity.

\subsection{K-L Eigenmodes as Probes of the Power Spectrum}
 
Figure 3 shows the expectation value of the ``power spectrum'' of the
K-L expansion for the CfA2 slice, in analogy to the familiar power per mode
of the Fourier expansion. 
When we expand the observations $\Dbf$ in the fluctuation eigenmodes,
the expectation value of the total
power per mode is the eigenvalue for that mode
plus a contribution from the mean density.
The total power per mode has components from
the clustering signal, noise, and mean density (where the primes
denote the whitened quantities),
\begin{eqnarray}
\langle B_n^2 \rangle & = & \langle (\Pbf_n^{\dag}\Nbf^{-1/2}\Dbf)
(\Pbf_n^{\dag}\Nbf^{-1/2}\Dbf)^{\dag}\rangle
\nonumber \\
  & = & \Pbf_n^{\dag} \langle \Dbf' \Dbf'^T \rangle \Pbf_n \nonumber \\
  & = & \Pbf_n^{\dag} \Rbf' \Pbf_n \nonumber + 
\Pbf_n^{\dag}\langle \Dbf'\rangle \langle \Dbf'\rangle \Pbf_n \\
  & = & \Pbf_n^{\dag} (\Sbf' + \Nbf' + \Ebf') \Pbf_n,
\end{eqnarray}
where $\Dbf'$ is the whitened vector of cell counts,
$\Dbf'=\Nbf^{-1/2} \Dbf$,
and using the separation of the whitened correlation matrix into the
sum of matrices, $\Rbf' = \Sbf' + \Nbf'$.
If shot noise is the only source of noise in the counts
($N_{ij}=n_i\delta_{ij}$), then these
matrices have the simple forms
\begin{eqnarray}
E'_{ij} & = & n_i^{1/2}n_j^{1/2}\nonumber \\
S'_{ij} & = & n_i^{1/2}n_j^{1/2} \xi_{ij} \\
N'_{ij} & = & \delta_{ij}.\nonumber
\end{eqnarray}

The contribution to the power per mode from the mean density is
\begin{eqnarray}
E_n^2 & = &  \Pbf_n^{\dag} \Ebf' \Pbf_n = \langle B_n \rangle ^2 \nonumber \\
      & = & \left (\sum_{i=1}^M \Psi_n(\xbf_i) n_i^{1/2} \right)^2 \nonumber \\
      & = & \left (\int d^3x\: F_n(\xbf) \bar{n}^{1/2}(\xbf) \right) ^2.
\end{eqnarray}
In the last line we take the limits $M\rightarrow\infty$ and 
$V_i\rightarrow 0$, and $F_n(\xbf)$ is defined in equation (25).
The power term $E_n^2$ quantifies the relative sensitivity of each mode to
the mean; if the eigenmode fluctuates around zero, then this integral
vanishes.
To eliminate any dependence on the mean density, we could exclude from
our analysis those few modes which do carry this information.

The variance of $B_n$ is the sum of the clustering and shot noise power,
\begin{equation}
\langle \delta B_n^2 \rangle = \langle B_n^2\rangle - \langle B_n\rangle^2
= S_n^2 + N_n^2.
\end{equation}
By construction,
the noise power per mode is constant,
\begin{equation}
N_n^2  = \Pbf_n^{\dag} \Ibf \Pbf_n = 1
\end{equation}

The clustering power per mode is
\begin{eqnarray}
S_n^2 
& = & \Pbf_n^{\dag} \Sbf' \Pbf_n  \nonumber \\
& = & \sum_{i=1}^M \sum_{j=1}^M \Psi_n^*(\xbf_i) \Psi_n(\xbf_j) 
n_i^{1/2} n_j^{1/2} \xi_{ij} \nonumber \\
& = & \int d^3x \int d^3x' \: F_n(\xbf) F_n(\xbf') \bar{n}^{1/2}(\xbf)
\bar{n}^{1/2}(\xbf') \xi(\xbf, \xbf')
\nonumber \\
& = & \int d^3x \int d^3x' \: G_n(\xbf) G_n(\xbf') \xi(\xbf, \xbf')
\nonumber \\
& = & \int {d^3k\over (2\pi)^3}\,  |\tilde{G}_n(\kbf)|^2 P(\kbf),
\end{eqnarray}
where we define the function
\begin{equation}
G_n(\xbf)=F_n(\xbf)\bar{n}^{1/2}(\xbf).
\end{equation}
The function $G_n(\xbf)$
has Fourier transform $\tilde{G}_n(\kbf)$, which is the convolution
of the Fourier transform 
of the eigenmode with the Fourier transform of the square root
of the selection function for the survey.
The last line in equation (32) 
follows by subsituting in the Fourier transform relation 
\begin{equation}
\xi(\xbf)={1\over (2\pi)^3}\int d^3k\: P(\kbf) e^{-i\kbf \cdot \xbf}.
\end{equation}
This Fourier transform relation does not hold exactly in redshift space,
because peculiar velocities couple real-space 
Fourier modes with different wavelength
(Zaroubi \& Hoffman 1994).
Therefore, this substitution is
only approximately correct, but serves to illustrate
how each K-L mode is a spatial filter that samples power from
the range of wavenumber described by its Fourier window 
$|\tilde{G}_n(\kbf)|^2$.
The clustering signal per mode is  the integral of the 
product (not the convolution) of this window with the power spectrum.

Figure 4 plots the two-dimensional Fourier windows,
$ |\tilde{G}_n(\kbf)|^2$, 
of the eigenmodes that are shown in Figure 1,
sampling only wavevectors in a plane close to the slice
(we use a two-dimensional Fourier transform for this example
because the CfA2 slice is narrow in the declination direction, and therefore
contains little information in the direction normal to the slice). 
In Figure 5 we plot the passbands in the Fourier domain for these same
eigenmodes. 
These passbands are the Fourier window functions
averaged over all directions $\hat{\kbf}$.

As we increase the volume covered by a survey, the Fourier windows of
its eigenmodes narrow and we probe the fluctuations with
increasing resolution.  
In the limit of an infinite volume, these windows approach
delta functions, and
the K-L eigenmodes become 
identical to the plane waves of the Fourier expansion,
modulated by weighting to account for the variation with distance of the 
selection function.

Because the spatial window function of a galaxy survey typically
has sharp angular limits,
the Fourier windows of its eigenmodes can have sidelobes due to
``ringing'' at the survey boundary.
It is tempting to design and apply	
a filter that smooths these edges (applied as a set of weights
to the galaxies) and thereby remove this ringing.
However, one does so at the cost of throwing away signal near the survey 
boundaries.
For the purpose of model testing, as described in the next section,
such a smoothing reduces the statistical
power of the data; the K-L expansion is a complete basis and every mode
is weighted in the likelihood function by its signal to noise ratio.
Appendix A shows that the K-L expansion is an optimal basis for testing
the likelihood of clustering models.
If we weight the data {\it a priori}, then we can only decrease the
discriminatory power of this statistic.

The K-L transform addresses the question of how to weight data in
different regions to yield an optimal estimator for very large
wavelength fluctuations and therefore 
is particularly useful for probing large-scale
clustering using surveys with very complex geometry.
Sensitivity to very large-scale density
fluctuations is limited by the total volume sampled which,
for a survey comprised of disjoint subregions, is the effective 
volume of the ``meta-survey," not that of an individual sub-region.
A limiting case is when the survey is a sum of delta functions centered
on randomly selected galaxies (this is the sparse sampling strategy espoused
by Kaiser 1986).
Surveys conducted with multifiber spectrographs typically populate 
the survey volume with an ensemble of pencil beams or slices.
Elsewhere (Szalay et al. 1993; Vogeley 1995) we
examine how the Fourier window function of such a survey 
depends on the window functions
of the individual subregions as well as their combination.
Individually, each of the subregions probes the fluctuation spectrum
with its own very broad window function (the volume is narrow, thus
the window function is broad), and this ``auto-correlation''
provides a poor estimate of large-scale clustering (in the sparse-sampling
limit, the auto-correlations are pure shot noise, with equal power
at all wavenumbers).
The window function for clustering power that arises from the contrast
in density between disjoint regions is significantly narrower, lacking
the sidelobes of the individual regions' geometries, 
and thus provides a cleaner probe of large-scale power.
One suggestion is to examine only this ``cross-correlation'' part
of the observed fluctuations.
However, such a power spectrum estimate is less than optimal because
it does not use any of the fluctuation signal within each subregion,
and is statistically more complex because the cross-correlation component
is not positive definite.

In the K-L transform,
the auto-correlation power and cross-correlation power are both represented
by the eigenmodes.
In this way we obtain the best possible resolution (from the cross-correlation
modes) without sacrificing statistical power (because the auto-correlation
modes still contribute to the likelihood).
In Figure 6 we display several of the eigenmodes for a survey
comprised of narrow beams within the limits of the CfA2 slice.
Comparison with the eigenmodes of the full slice (Fig. 1) shows that
the modes of the partial survey sample the large-scale fluctuations
(compare modes 1-6 in Fig. 6 with modes 1, 2, 4, 7, 3, and 5  respectively,
in Fig. 1), as well as the fluctuations within the individual beams.

\section{Statistical Tests of Cosmological Models}

After we derive the K-L eigenmodes and apply this transform
to the observations, the next step is to use the observed K-L coefficients
to test cosmological models.
We test these models' predictions of the mean and second moments 
(e.g., the power spectrum)
of the galaxy density field, which are sufficient to predict 
the mean and covariance matrix of the K-L coefficients.
To find the best fit model and the corresponding confidence region of model
parameters, we apply Bayesian methods to compute the
posterior probability that a model would yield the observed expansion
coefficients.
This probability is 
proportional to the likelihood of the observed set of coefficients,
using the covariance matrix for each model. 
This approach contrasts with standard methods for power spectrum
estimation, 
in which one averages the power per mode in bins of wavenumber,
and tests the likelihood of the observed power using the covariance
matrix of the power that the model predicts.

\subsection{Bayesian Model Testing}

Our use of prior knowledge of the selection function of the survey and the
noise and clustering properties of the galaxy density field
gives a Bayesian flavor to our
method for finding the eigenmodes of the survey.  We continue in this
spirit and follow Bayesian methods to find the model most likely to
have yielded the data set that we observed.  In this section we
outline the procedure for estimating the posterior probability of a
model, with particular attention to how use of the K-L transform
simplifies many of the necessary steps and how a Bayesian approach
increases the discriminatory power of the statistics.  Though our
particular interest is in estimating the power spectrum, we describe
how to simultaneously estimate the mean density, power spectrum of
density fluctuations, and redshift distortion parameters from the
observations.

It might seem that the model assumed for construction of the K-L eigenmodes
would prejudice the likelihood analysis of other models.  
This is not the case; the K-L eigenmodes are a complete orthonormal basis
regardless of the assumed power spectrum.
We allow the mean density to vary as a parameter of the models, so that
the choice of mean does not bias the estimate of large wavelength fluctuations.
However, a bad initial ``guess''
makes the subsequent analysis slightly more difficult because,
although the K-L modes would still be orthonormal, they would not quite
be statistically orthogonal.
Changing $P(k)$ changes
the expected signal to noise in a pixel, thus the weighting for probing a
given range of wavelength scales will not be optimal.  Therefore, a
bad initial guess at $P(k)$ causes covariance and less than optimal
signal to noise of the eigenmodes, which
broaden the confidence regions, but does not change the peak 
of the likelihood function. 
If our initial guess turns out to be
particularly bad,
we can iterate the estimation procedure by using the best fit model
to construct better eigenmodes.

To test proposed models we employ Bayes' theorem to compute the
posterior probability of a set of model parameters  (for an excellent
review of application of Bayesian methods to astronomical problems, 
see Loredo 1990 and references therein).  Given a set of
observations $D$ and additional ({\it prior}) information $I$, the
posterior probability of a model that is specified by the
parameters $\{\theta_i\}$ is
\begin{equation}
P(\{\theta_i\} | DI)=
P(\{\theta_i\} | I) { P(D |\{ \theta_i\} I) \over P(D | I) }.
\end{equation}
The first factor is the Bayesian {\it
prior}, which encodes our other information (i.e., apart from the new
data) about the probability of this model being correct.  The
likelihood function (as we usually think of it) enters into 
$P(D |\{\theta_i\} I)$.  
The denominator is essentially a normalization
constant, determined by requiring that the posterior probability integrate
to unity over the entire parameter space of possible models.

In our case, the observations $D$ are the observed galaxy counts, 
represented by the K-L coefficients,
and the model parameters $\{\theta_i\}$ describe the expectation
value and second moments of the galaxy density in redshift space.  These
parameters may include the mean density of galaxies, the selection
function for the survey, the power spectrum of galaxy density
fluctuations, and the parameters that describe the distortion of
redshift space due to peculiar velocities.
The prior information $I$ might include, e.g., estimates of the 
mean density, $\Omega^{0.6}/b$, or the power spectrum, obtained from
other surveys or methods, as well as obvious constraints such as 
$\Omega>0$ and $P(k)>0$.

\subsection{Bayesian Priors}

The prior distribution $P(\{\theta_i\} | I)$ describes the probability
distribution of the model parameters in the absence of the new data
$D$.  By adopting an {\it informative} prior probability for the
parameters, we narrow the confidence region of the posterior
probability for the model, yielding better constraints on the
parameters of interest.  An example is the mean galaxy density: most
often, we estimate the mean density from the data themselves, but this
estimate may differ from the cosmic mean due to shot noise and
clustering of galaxies on the scale of the survey volume.  Rather than
choose a single best value for the mean density and risk
underestimating the power spectrum on large scales (see section 1.3 for
a discussion of this problem), we can include both the mean density
and the power spectrum as parameters of the model and simultaneously
fit for both. A prior probability distribution that describes the most
likely value for the mean galaxy density and its uncertainty will
narrow the range of acceptable power spectra.  Another example is
redshift distortions: they alter the shape of the power spectrum
in such a way that it may be
difficult to differentiate between a model that has a great deal of
power on large scales in real space, but low $\Omega$ and thus small
redshift distortions, and a model with intrinsically moderate
large-scale power but which has large streaming velocities.  
If we know, from other observations, that $\Omega^{0.6}/b=1\pm 0.3$ and that
the uncertainty is normally distributed, then we can include this
probability distribution in the prior to better constrain the real-space
power spectrum.

Of course, we always have the option to ignore everything else that
we know about the universe and adopt a uniform prior, i.e.,
equal prior probability for all values of a parameter (or almost all --
for example, we still want to constrain $\Omega >0$ and $P(k)>0$).

Note that these Bayesian priors differ from the `prior model' that we used
to construct the eigenmodes, though we would be wise to construct the
eigenmodes using model parameters that we deem {\it a priori} most
likely.

\subsection{Likelihood Functions}

The second factor in the posterior probability of a particular model
(eq. [35]) is 
the {\it likelihood} that
these data deviate from their expectation values as widely as observed.
We observe the counts $\Dbf$, whiten these observations with $\Nbf^{-1/2}$,
and expand the whitened counts in the eigenvectors $\Pbf_n$ to obtain
the K-L coefficients $\Bbf$.
Each model predicts the mean $\langle \Bbf \rangle_{\mathrm{model}}$ and
covariance matrix ${\bf C}_{\mathrm{model}}$ of these coefficients.
The covariance matrix ${\bf C}_{\mathrm{model}}$ has elements
\begin{eqnarray}
C_{ij}
& = & \langle (B_i-\langle B_i\rangle) (B_j-\langle B_j\rangle)
 \rangle \nonumber \\
& = & \Pbf_i^{\dag} \Rbf'_{\mathrm{model}} \Pbf_j  \nonumber \\
& = & \Pbf_i^{\dag}\Nbf^{-1/2} \Rbf_{\mathrm{model}}  \Nbf^{-1/2}\Pbf_j,
\end{eqnarray}
where we compute $\Rbf_{\mathrm{model}}$
by substituting the model's selection function, 
power spectrum, etc., into equation (21).

The quadratic form
\begin{equation}
\chi^2 = (\Bbf - \langle \Bbf \rangle_{\mathrm{model}})^T 
{\bf C}^{-1}_{\mathrm{model}} (\Bbf - \langle \Bbf\rangle_{\mathrm{model}})
\end{equation}
measures the goodness of fit of the model.
For the set of model parameters used to derive the eigenmodes,
$C_{ij}=\delta_{ij}\lambda_i$, where $\lambda_i$ is the eigenvalue of the
$i^{th}$ mode. In this case, $\chi^2$ is simply the number of degrees
of freedom, i.e., the number of eigenmodes.
In general, because 
the model under consideration differs from the model used to
construct the eigenmodes, ${\bf C}_{\mathrm{model}}$ 
will not be diagonal because
the eigenmodes of the prior model are not exactly statistically orthogonal
under the hypothesis being tested.
Although repeated inversion of ${\bf C}_{\mathrm{model}}$ would seem to be
computationally intensive,
in practice the models that we will test differ most greatly on large
wavelength scales (where our current knowledge is most uncertain), 
in which case the off-diagonal elements of
${\bf C}_{\mathrm{model}}$ would be isolated in one corner of the
matrix and so these matrices
should quickly diagonalize to within machine precision.

To compute the probability $P(D | \{\theta_i\} I)$ for a specific
model we must specify the probability distribution of $\chi^2$.  If
the observed density field and therefore the probability distribution
of the K-L coefficients is Gaussian, then the likelihood function of
the data is a multivariate Gaussian,
\begin{equation}
\like(\Bbf \mid \mathrm{model})= (2\pi)^{-M/2} 
|\det {\bf C}|^{-1/2}
\exp ( - \chi^2 / 2),
\end{equation}
where $M$ is the number of eigenmodes included in $\chi^2$.

Why does $I$ appear in $P(D | \{\theta_i\} I)$?  One reason is that
we make certain assumptions that apply to the models in consideration.
For example, we think it reasonable to use a
Gaussian likelihood function to probe large wavelength scales only 
because some
previous observations and prevailing theoretical prejudice tell us
that the distribution of matter on large wavelength scales is nearly
Gaussian.  
Such a separation of linear and non-linear scales is possible because the
K-L eigenmodes isolate the statistically independent bands of power
on different scales.
It is important to remember that such an assumption
implicitly sets the family of models that we test.

A more accurate computation of the likelihood, and one that
would allow inclusion of modes that sample smaller wavelength scales,
would use a non-Gaussian likelihood function.  Amendola (1994)
describes how to construct such a function using the Edgeworth
expansion, where the observed higher order clustering properties are
used to compute the broadening of the probability distribution of
$\chi^2$ due to non-linear clustering of galaxies.
Of course, such an extension to non-linear scales requires prediction
of these higher order moments for the model in question.

\subsection{Computing the Confidence Region}

To find the confidence region of likely model parameters,
we employ equation (35)
to compute the posterior probability for models that span
the available parameter space,
locate contours of constant posterior probability, and then integrate
the probability within the contours.
The integral of probability density within these contours yields the Bayesian
confidence regions.
The assignment of absolute probability to a model,
as opposed to the relative likelihood of different models,
requires that we know the denominator $P(D|I)$ of equation (35), or that
the tested parameters space includes all possible models,
in which case we determine this normalization by integrating the posterior
probability over the entire parameter space.

Bayesian and frequentist methods of analysis differ in their interpretation
of confidence regions. In the Bayesian view, the confidence
region that we obtain is the region of parameter space in which the
models have the same posterior probability given the observed data
and prior information.
In contrast, the confidence region obtained in a frequentist
analysis describes the distribution of estimated parameters that we
could expect to measure for a population of similar data sets, if the true
parameters are those which best fit the observed data.

To determine the probability density of some subset of the model
parameters, we can marginalize (integrate eq. [35]) over
the distribution of all other parameters.
For example, to find the most likely power spectrum parameters, we could
marginalize over the distribution of the mean density, or vice versa.

At this point we can assess whether the model used to construct the
eigenmodes was a good initial guess, by examining where this prior model
lies in the confidence region of tested models.

\subsection{``Model Independent'' Plots of the Estimated Power Spectrum?}

Often the first result that we desire from a power spectrum analysis
is a plot of the estimated power spectrum and error bars at several
wavenumbers.
We can produce such a plot if we apply the methods above to test
a model in which the power spectrum parameters are simply the average power in 
bins of $\delta k$.
The most likely value and 68\% confidence region of the marginal 
distribution of the power in each bin of wavenumber yield the 
quantities that we want to plot.
Very wide bins will yield small uncertainty in the power per bin, because many
eigenmodes contribute to the total power, but poor resolution of
features in the power spectrum. 
Narrower bins would better show such features, but the uncertainty
in power per bin would be correspondingly larger.

A problem with such plots is that they do not communicate the covariance
of power estimates in different bins.
Only when 
the bins of wavenumber are wide compared to the Fourier windows of the
K-L eigenmodes  
can we approximate the power in different bins as being independent. 
The tradeoff between resolution and uncertainty that one faces in producing
such a plot (cf. Tegmark 1995 for
discussion of the ambiguity between ``vertical'' and ``horizontal''
error bars on the power spectrum)
is, however, one of graphical semantics rather than
science. Formal testing of proposed models should use the full likelihood
function of the observed coefficients 
rather than ``chi-by-eye'' comparisons of the
estimated power spectrum and uncertainties with different model power
spectra.
The K-L modes are the
narrowest possible statistically orthogonal linear combinations of the
Fourier modes.  
A larger number of spatial filters would not be
independent; a lesser number would cause loss of statistical power and
resolution.
Following the Bayesian method that we advocate,
no binning is required and the modes with large signal to
noise naturally receive higher weights in the likelihood function.
The tradeoff between resolution and uncertainty enters when we select
the models that we test (number of parameters in the $P(k)$ model)
rather than in binning the power estimates.

Another problem with ``model independent'' plots of the power spectrum
is that, as we note in section 3.2, 
we require specification of a cosmological model in order
to compute comoving coordinate distances from the observed redshifts.

\section{Discussion}

In this paper we describe a transform method for analyzing galaxy
redshift surveys that
allows estimation of the mean density, power spectrum,
and redshift-space distortions, 
and which may prove useful for other purposes in characterizing other
properties of the observed large-scale structure. 
In summary,
the basic steps in this analysis are:
\begin{itemize}
	\item Divide the survey volume into cells.
	\item Select a prior model and compute the correlation matrix of the
galaxy density fluctuations predicted by this model.
	\item Find the eigenvectors and eigenvalues of the correlation matrix,
after applying a whitening transformation.
	\item Bin the observed galaxy counts into cells and compute 
the K-L transform of these observations.
	\item Select the cosmological models to be tested and the prior 
information to be included in the model testing. 
	\item Compute the posterior probability for 
a range of values of the model parameters and find the confidence
regions of these parameters.
	\item If necessary, use the best fit model to construct a more
optimal set of eigenmodes and iterate.
\end{itemize}

\subsection{Comparison with Direct Fourier Analysis}

Because
we use the K-L transform rather than the Fourier transform to decompose
the observations, and work with the probabilities of the coefficients
themselves rather than the power per mode,
we gain several advantages over the standard method described in
the Introduction.
The K-L modes are, by construction, statistically orthogonal,
thus we begin the likelihood analysis with exactly the required number
of probes of clustering power
(one could find the linear combinations of Fourier modes that
are statistically orthogonal, but
such a procedure is just another route to the K-L eigenmodes,
where the eigenvectors are a set of weights applied to the Fourier
modes rather than to the observed cell counts, and would require the
additional step
of first computing the Fourier transform of the data).
By testing the likelihood
of the observed K-L coefficients rather than the
probability distribution of the power per mode, we require
lower order assumptions about
the probability distribution of the galaxy density.
We treat the mean density as just another parameter of the model being
tested and simultaneously fit both the mean density and power spectrum,
thereby differentiating between the
case in which the observed mean density is equivalent to the cosmic
mean, and the case where the observed mean differs from the cosmic
mean due to density fluctuations on the scale of the survey.
The Bayesian approach provides a clear method for better constraining the
confidence region of power spectrum parameters
by inclusion of prior knowledge of the mean density in the
calculation of the posterior probability of a model.

As Appendix A demonstrates,
the K-L transform method yields the optimal weighting for testing 
the likelihood of clustering models.
This is a more general result than
previous derivations of minimum variance weighting (e.g., FKP),
which find the optimal weighting for Fourier analysis under specific
idealized conditions.
Tailoring of the weighting scheme to obtain better resolution in the
power spectrum estimator (Tegmark 1995)
achieves this improvement at the expense of statistical power to
differentiate between clustering models.

\subsection{Comparison with Other Transform Methods}

The K-L expansion
is the optimal basis set if we truncate the expansion to the subset
of modes with highest signal to noise, and therefore
may be considered a
form of Principal Component Analysis (PCA) or factor analysis.
PCA more commonly (but not exclusively)
describes the case in which we
reduce a set of observed variables to a smaller number of linear
combinations which completely describe the observations. 
Factor analysis seeks a lower-dimensional representation that describes
all of the correlations among the data.
Because there is non-zero clustering power on all scales,
all of the K-L eigenmodes are required to provide a
complete representation of the observed galaxy density field.

Because the K-L eigenmodes are eigenvectors of the correlation matrix,
they differ from basis functions that, although orthonormal, are not
statistically orthogonal.  
We show that this representation of the redshift-space density field
is optimal for testing models for the power spectrum for any survey.
Other transform methods offer advantages for other analyses.
For example,
Fisher et al. (1994) expand the observed redshift-space density field
in spherical Bessel functions and spherical harmonics 
(Fourier-Bessel expansion) in order to reconstruct
the real-space density, velocity field, and potential via a Wiener filtering
method.
Similarly, Heavens \& Taylor (1995) 
advocate use of this same representation for
estimating the redshift-space anisotropy of galaxy clustering.
The Fourier-Bessel expansion simplifies computation of the
effect of redshift distortions on the galaxy density field because
these are eigenfunctions of the Laplacian in Poisson's equation, which
describe redshift distortions in linear theory.
For an all-sky survey, redshift space distortions only couple different
radial modes, thus the transformation from real to redshift space is
straightforward.
The advantages of this approach are lost, however, if there are
large gaps in the sky coverage of the survey, which destroy the
orthonormality of the spherical harmonics.
The requirement that one truncate the expansion to a finite number of
modes is analogous to our division of space into a finite number of pixels.
These functions are not statistically orthogonal, thus model testing
using this expansion requires inversion of generally very non-diagonal
matrices.

In parallel with the development of alternative means of power
spectrum estimation from galaxy surveys, various transform methods
have also been developed for estimating the power spectrum at
recombination from the COBE DMR maps.  
Analysis of the CMB anisotropy faces many of the problems posed by
analysis of the redshift-space galaxy distribution, with the
simplifications that the mean is extremely well determined,
redshift distortions (apart from our own dipole
motion with respect to the CMB) play no role, the window
function on the sky is relatively simple, the fluctuations are
Gaussian (in nearly all theories under consideration), and the noise
per pixel is almost constant and nearly uncorrelated.

The ``signal-to-noise eigenmodes'' derived by Bond (1994) for analysis
of the COBE DMR maps are constructed in similar fashion to the eigenmodes of
the galaxy redshift distribution that we describe in this paper.
Both methods apply the Karhunen-Lo\`eve transform to represent the
observations.
In the case of the COBE DMR analysis, one
uses an assumed model for the power spectrum at recombination,
as well as the correlation matrix of the pixel-pixel noise (in this case
the noise correlation matrix depends on the instrument, in contrast
to the dependence on the sampling density of galaxies), to construct
a complete orthonormal basis in which both the signal and noise correlation
matrices of the expansion coefficients are diagonal.
Expansion of the pixel maps in this basis allows straightforward Bayesian
model testing.
Because the modes are ordered by signal to noise ratio,
this basis set is also optimal for Wiener filtering of the pixel maps.

An example of a model independent method (in the sense of requiring no
{\it a priori} clustering model) is that of G\'orski (1994), who
derives a set of functions that, unlike the spherical harmonics, are
orthonormal over the cut sky (the areal region left after removing
areas close to the Galactic plane) of the COBE DMR maps.  This
procedure for Gram-Schmidt orthogonalization yields linear
combinations of spherical harmonics that are both orthonormal and that
are as compact as possible in the spherical harmonic function space.
After removing the monopole and dipole contributions (which, unlike
the case of galaxy observations, are sufficiently well determined to
justify their removal), Bayesian analysis of the expansion
coefficients yields estimates of the power spectrum parameters.
Neither the noise nor the signal correlation matrices are diagonalized
by this transformation.  As we see above, such a diagonalization
requires assumption of a prior model to form the eigenmodes, which
this method seeks to avoid.

\subsection{Other Uses of the K-L Transform}

The K-L eigenmodes form a complete basis for representing the observations,
with no loss of phase information, and thus may be useful for
analyses other than power spectrum estimation.
One such application is smoothing of the density field by removal
or suppression of modes that sample short wavelength scales.
This form of smoothing could be used for studies of the morphology
and topology of large-scale structure and for identification of
superclusters in sparse data.
Another application is 
optimal reconstruction of the galaxy density field, facilitated
by the signal to noise properties of the eigenmodes (cf. Bond 1994
and Fisher et al. 1994 regarding Wiener filtering for reconstruction
of the CMB anisotropy and the galaxy density field, respectively).
When applied to spectroscopic observations of galaxies, the K-L
transform yields an elegant means of spectral classification
(Connolly et al. 1995).

\subsection{Applications to Galaxy Survey Data}

This paper is the first in a series in which we apply eigenmode
analysis to a variety of data sets.  These include the pencil beam
redshift surveys, early redshift
observations from the Sloan Digital Sky Survey, and the spatial
distribution of quasar absorption line systems.  For future surveys
(including planning for the SDSS), we can design the optimal geometry
and sampling for the available survey resources, using the K-L
transform as a method for estimation of the uncertainty in $P(k)$ for
an arbitrary survey.  Ultimately, we hope to use this transform method
to combine all of the available galaxy redshift data and thereby
obtain the best possible measurement of the power spectrum of galaxy
density fluctuations.

\acknowledgements

We thank the referee, John Peacock, for helpful comments on the text.
We acknowledge support from NSF grant AST-9020380 and
the Seaver Foundation.
Partial support for this work was also provided by NASA through grant
number AR-05813.01-94A from the Space Telescope Science Institute,
which is operated by the Association of Universities for Research
in Astronomy, Inc., under NASA contract NAS5-26555.

\appendix
\section{An Optimal Basis for Model Testing}

To most strongly differentiate among clustering models, we
want the volume of the confidence region of the parameter estimates
to be as small as possible.
In other words,
we want the likelihood function $\like(\Bbf|{\mathrm{model}})$
to be sharply peaked at the true model.
We accomplish this by expanding the observations $\Dbf$ in a set of
basis functions for which the likelihood of the coefficients $\Bbf$
decreases as steeply as possible as we perturb the clustering model
from the best fit model.

The maximum likelihood estimate of the model parameters
$\{\theta_i\}$ occurs when the gradient of the likelihood function
(otherwise known as the score function) is set to zero.
Here we assume a Gaussian likelihood function,
\begin{eqnarray}
\like(\Bbf | {\mathrm{model}})
& = & (2\pi)^{-1/2}(\det \Cbf)^{-1/2}
\exp\left[
-(\Bbf - \langle \Bbf\rangle)^T \Cbf^{-1}(\Bbf - \langle \Bbf \rangle)/2
\right]
\\
& = & (2\pi)^{-1/2} (\det \Cbf)^{-1/2} \exp\left [
-\trace(\Cbf^{-1}\Zbf)\right],
\end{eqnarray} 
where $\Bbf$ are the expansion coefficients in some 
basis $\{\Pbf_n\}$, $\Cbf$ is the
covariance matrix of these coefficients, and $\Zbf$ is the sample
covariance matrix
\begin{equation}
\Zbf=(\Bbf - \langle \Bbf\rangle)(\Bbf - \langle \Bbf \rangle)^T.
\end{equation}
The maximum of $\like$ is also the minimum of $-2\ln \like$,
so we evaluate the derivative of the log-likelihood function with respect
to the one of the model parameters $\theta_i$:
\begin{eqnarray}
{\partial \over \partial \theta_i}(-2 \ln \like) & = &
{\partial \over \partial \theta_i} \left\{
\ln(\det\Cbf)+\trace (\Cbf^{-1}\Zbf) \right\} \\
& = & \trace [\Abf] - \trace \left[
\Abf\Cbf^{-1}\Zbf \right] 
+\trace \left[
\Cbf^{-1} {\partial \Zbf \over \partial \theta_i}\right],\nonumber
\end{eqnarray}
where we define
\begin{equation}
\Abf_i=\Cbf^{-1}{\partial \Cbf \over \partial \theta_i}.
\end{equation}
Setting this derivative (eq. [A4]) to zero we obtain the maximum 
likelihood estimate of $\{\theta_i\}$.

The optimal basis functions are those which yield the narrowest
confidence region around the best fit model $\{\hat{\theta}_i\}$.
The volume of the confidence region depends on the Fisher information
matrix, or second gradient of the log-likelihood function, which measures
how steeply the likelihood falls as we move away from the best fit model,
evaluated at the position of the best fit model.
The second derivatives of the log-likelihood function are
\begin{eqnarray}
{\partial^2 \over \partial \theta_i \partial \theta_j}(-2 \ln\like) & = &
\trace \left[ {\partial \Abf_j \over \partial \theta_i} \right]
- \trace \left[ {\partial \Abf_j \over \partial \theta_i}
\Cbf^{-1}\Zbf \right]
+ \trace \left[ \Abf_j \Abf_i \Cbf^{-1}\Zbf \right ] \\
& - & \trace \left [ \Abf_j\Cbf^{-1}
{\partial \Zbf \over \partial\theta_i }\right ]
- \trace\left[ \Abf_i \Cbf^{-1} {\partial \Zbf \over\partial\theta_j}
\right]
+ \trace\left[\Cbf^{-1}{\partial^2 \Zbf \over \partial \theta_i
\partial \theta_j}
\right].
\nonumber
\end{eqnarray}
To evaluate the derivatives, we need the expectation value of the
sample covariance matrix and its derivatives:
\begin{eqnarray}
\langle \Zbf \rangle & = & \Cbf \\
\left \langle {\partial \Zbf \over \partial \theta_i}\right \rangle & =&  0 \\
\left \langle {\partial^2 \Zbf \over \partial \theta_i \partial\theta_j}
\right \rangle
& =& {\partial \langle\Bbf\rangle \over \partial \theta_i}
{\partial \langle\Bbf\rangle ^T \over \partial \theta_j}
+ {\partial \langle\Bbf\rangle \over \partial \theta_j}
{\partial \langle\Bbf\rangle ^T \over \partial \theta_i}
\end{eqnarray}
Using these identities we obtain
\begin{equation}
{\partial^2 \over \partial\theta_i\partial\theta_j} (-2\ln\like)
= \trace\left[ \Abf_i\Abf_j\right]
+2{\partial \langle\Bbf\rangle^T \over \partial\theta_i} \Cbf^{-1}
{\partial \langle\Bbf\rangle^T \over \partial\theta_j}.
\end{equation}

To optimize the basis functions 
for sensitivity to the clustering model, let us assume
that we know the true mean density, thus
\begin{equation}
{\partial^2 \over \partial\theta_i\partial\theta_j}(-2 \ln\like)
= \trace\left[ \Abf_i\Abf_j\right]\\
= \trace\left[ \Cbf^{-1} {\partial\Cbf \over \partial \theta_i}
\Cbf^{-1}{\partial \Cbf \over \partial\theta_j} \right].
\end{equation}
The covariance matrix is the sum of signal and noise components
\begin{equation}
\Cbf= \Pbf^{\dag}\Sbf\Pbf + \Pbf^{\dag}\Nbf\Pbf,
\end{equation}
where $\Sbf$ and $\Nbf$ are the signal and noise correlation matrices
of the observations.

If we perturb the clustering model from its best fit value 
$\Sbf(\{\hat{\theta}\})$, we
vary the signal correlation matrix as
\begin{equation}
\Sbf = \hat{\Sbf} + {\bf p}\hat{\Sbf},
\end{equation}
where ${\bf p}$ is the matrix describing the pertubation.
We want to minimize
the second gradient of the likelihood with respect to this pertubation,
or maximize
\begin{equation}
{\partial^2 \over \partial {\bf p}^2} (-2\ln\like)
= \trace\left[
\Cbf^{-1} {\partial \Cbf \over \partial {\bf p}}
\Cbf^{-1} {\partial \Cbf \over \partial {\bf p}}
\right ].
\end{equation}
The argument of $\trace[...]$ is the product of two identical
symmetric matrices, thus the maximum of equation (A14) is also the maximum 
of
\begin{eqnarray}
\trace \left \{ \Cbf^{-1}{\partial \Cbf\over \partial {\bf p}}
\right \}
& = & \trace \left \{
\left[ \Pbf^{\dag}(\Sbf+\Nbf)\Pbf\right]^{-1}
{\partial \over \partial {\bf p}} \Pbf^{\dag}(\Sbf+\Nbf)\Pbf
\right \} \\
& = & \trace \left \{
\left[\Pbf^{\dag}(\Sbf+\Nbf)\Pbf\right ]^{-1} \Pbf^{\dag}2\Sbf\Pbf 
\right \}
\nonumber\\
& = & 2\trace \left \{ \Pbf^{\dag}(\Sbf+\Nbf)^{-1} \Sbf\Pbf \right \}.
\nonumber
\end{eqnarray}

To find the optimal basis functions, we now maximize equation (A15)
with respect to changes in the basis functions $\{\Pbf_n\}$,
which are columns of the transformation $\Pbf$,
subject to the constraints that these functions be orthonormal.
We solve this problem using the method
of Lagrange multipliers, and maximize the Lagrangian (not the likelihood,
though they momentarily share the notation $\like$),
\begin{equation}
\like = \Pbf^{\dag}(\Sbf+\Nbf)^{-1}\Sbf\Pbf+\Lambda(1-\Pbf^{\dag}\Pbf),
\end{equation}
where $\Lambda$ is the diagonal matrix of Lagrange multipliers.
We compute the gradient with respect to the matrix $\Pbf$ and set this
to zero,
\begin{eqnarray}
{\partial \like \over \partial \Pbf}
= (\Sbf+\Nbf)^{-1}\Sbf\Pbf - \Lambda\Pbf = 0
\end{eqnarray}
The optimal basis vectors are the solution to this
generalized eigenvalue problem, which we can 
rearrange to form a simple eigenvalue problem:
\begin{eqnarray}
(\Sbf+\Nbf)^{-1}\Sbf\Pbf&=&\Lambda\Pbf \nonumber\\
                \Sbf\Pbf&=&\Lambda\Sbf\Pbf+\Lambda\Nbf\Pbf \nonumber\\
                \Sbf\Pbf&=&\Lambda(\Ibf-\Lambda)^{-1}\Nbf\Pbf \nonumber\\
                \Sbf\Pbf&=&\Lambda\Nbf\Pbf\nonumber \\
\Nbf^{-1/2}\Sbf\Nbf^{-1/2}\Pbf &=&\Lambda \Pbf
\end{eqnarray}
where
$\Ibf$ is the identity matrix, $\Lambda(\Ibf-\Lambda)^{-1}\rightarrow\Lambda$
is justified because $\Lambda$ is an as yet unknown diagonal matrix,
and the elements of $\Nbf^{-1/2}$ are the square roots of $\Nbf^{-1}$.
$\Nbf^{-1/2}$ is a whitening transformation, which diagonalizes the
noise component of the covariance matrix.
Written in this fashion, the optimal functions for expanding the observations
are $\Pbf_n\Nbf^{-1/2}$, the product of a whitening transformation with
the eigenvectors
of the whitened signal correlation matrix of the best fit model.
We obtain the identical eigenvectors from 
the whitened covariance matrix,
\begin{equation}
\Nbf^{-1/2}\Rbf\Nbf^{-1/2}=\Nbf^{-1/2}(\Sbf+\Nbf)\Nbf^{-1/2}
=\Nbf^{-1/2}\Sbf\Nbf^{-1/2}+\Ibf.
\end{equation}

\setcounter{equation}{0}
\section{An Approximate Method for Computing $\xi_{ij}$}

Here we derive an approximation to the integral average of the correlation
function between two cells, $\xi_{ij}$.
Computation of this integral is complicated by redshift distortions, 
through which
the simple theoretical quantity in real space, $\xi(r=|\xbf-\xbf'|)$,
becomes a function of both the direction of $\xbf-\xbf'$ and the
distance of this pair from the observer.
Below we present a derivation that ignores redshift distortions.
This treatment is
sufficient to estimate the properties of the angle-averaged
redshift-space power spectrum, but falls short of using
the full statistical power of the eigenmode method.

By explicitly including the redshift-space distortions in the 
clustering model, we can simultaneously estimate both the real-space
power spectrum and the redshift-space distortions.
To do so we generalize the approximation described below,
using techniques
similar to those employed in Reg\H{o}s \& Szalay 1994.
Discussion of the physical motivation and details of our method
for modelling the redshift distortions is sufficiently
lengthy that we will include it in a future paper in this series.
The ``far-field'' approximation (which assumes that the directions
$\hat{\xbf}$ and $\hat{\xbf}'$ are parallel, thus 
$\xi(\xbf,\xbf')=\xi(\xbf-\xbf')$)
that is commonly used for examining
redshift distortions is not necessarily valid over the entire region
of the survey volume.
Our method does not rely on the ``far-field'' approximation and therefore
is accurate for any survey geometry.

We assume spherical geometry, where each cell is bounded by
\begin{eqnarray}
\varphi_l  <  \varphi\leq\varphi_u \nonumber \\
\vartheta_l  <  \vartheta\leq\vartheta_u \\
r_l  <  r<r_u \nonumber
\end{eqnarray}
where $\varphi,\vartheta,r$ are the normal Euler coordinates.
The central mass of each cell $i$ ($i = 1,2$) is $\bx_i$.
The distance between the two cells is $R=\vert \bx_2-\bx_1\vert$.
The vector $\bs_i$ points from 
the central mass of the $i$th cell to a position within that cell,
thus the position of a point in cell $i$ is $\br_i=\bx_i+\bs_i$.
We also define the vector $\bx = \bx_2-\bx_1$ between the
central masses of the two cells, as well as $\bs=\bs_2-\bs_1$
and $\br = \br_2-\br_1$.

We want to calculate the expectation value of the correlation function
between two cells ($i=1,2$)
\begin{equation}
\xi_{ij}  = 
{1\over V_1 V_2}\int d^3 \bs_1 \int d^3\bs_2\, \xi(r),
\end{equation}
where $\xi(r)$ is the theoretical correlation function, and $r=
\vert (\bx_2+\bs_2)-(\bx_1+\bs_1)\vert = \vert \br_2-\br_1\vert$.

Assuming that the distance between the cells is much larger than the 
size of each cell, we can expand $\xi(r)$ in a Taylor series up to
second order (as we find below, the first order cancels out, so
we do need the second order term):
\begin{equation}
\xi(r) = \xi(R)+\sum\limits_\alpha s_\alpha
\left({\partial \xi(r)\over\partial s_\alpha}\right)_R+
{1\over2}\sum\limits_{\alpha,\beta}s_\alpha s_\beta
\left({\partial^2 \xi(r)\over\partial s_\alpha \partial s_\beta}\right)_R
+\dots
\end{equation}
The first derivatives are
\begin{eqnarray}
{\partial\xi(r)\over\partial s_\alpha} & = &
{\partial\xi(r)\over \partial r}{\partial r\over\partial s_\alpha}=
\xi^\prime(r) {\partial \sqrt{R^2+\bs^2+2\bx\bs}\over\partial s_\alpha}
\\
& = & \xi^\prime(r) {1\over r}\left(s_\alpha  +x_\alpha\right) =
\xi^\prime(r){r_\alpha\over r}.\nonumber 
\end{eqnarray}
Expanding around $r = R$,
\begin{equation}
\left({\partial \xi(r)\over\partial s_\alpha}\right)_R=
\xi^\prime(R){x_\alpha\over R}.
\end{equation}
The second derivatives are
\begin{eqnarray}
{\partial^2\xi(r)\over\partial s_\alpha\partial_\beta} & = &
{\partial\over\partial s_\beta}\left(\xi^\prime(r){r_\alpha\over r}\right)
\\
& = & {\partial\xi^\prime\over s_\beta}{r_\alpha\over r}+
\xi^\prime(r) {1\over r}{\partial r_\alpha\over s_\beta} -
\xi^\prime(r) {1\over r^2} r_\alpha {\partial r\over s_\beta}\nonumber\\
& = &
\xi^{\prime\prime}(r){r_\beta\over r} {r_\alpha\over r} +
\xi^\prime(r) {\delta_{\alpha\beta}\over r} -
\xi^\prime(r) {r_\alpha\over r^2} {r_\beta\over r}.
\nonumber
\end{eqnarray}
Expanding around $r=R$ we obtain
\begin{equation}
\left({\partial^2 \xi(r)\over\partial s_\alpha \partial s_\beta}\right)_R=
\left(\xi^{\prime\prime}(R)-{1\over R}\xi^\prime(R)\right)
{x_\alpha x_\beta\over R^2}+
{\xi^\prime(R)\over R}\delta_{\alpha\beta},
\end{equation}
where $\delta_{\alpha\beta}$ is the Kronecker symbol.
Thus, the second order Taylor series expansion is
\begin{equation}
\xi(r) = \xi(R) + {\xi^\prime(R)\over R}\sum\limits_\alpha
s_\alpha x_\alpha + 
{1\over2}\left(\xi^{\prime\prime}(R) - {\xi^\prime(R)\over R}\right){1\over
R^2}\sum\limits_{\alpha,\beta} s_\alpha s_\beta x_\alpha x_\beta +
{1\over 2} {\xi^\prime(R)\over R}\sum\limits_\alpha s_\alpha^2.
\end{equation}

Next we evaluate the different orders of the approximation
\begin{equation}
\xi_{12}\approx \xi_{12}^{(0)} + \xi_{12}^{(1)} + \xi_{12}^{(2)}
\end{equation}
by integrating equation (B8) over the pair of cells:
\begin{eqnarray}
\avg{\xi_{12}}^{(0)} & = & {1\over V_1 V_2}\int\int d^3\bs_1 d^3\bs_2\,
\xi(R) = \xi(R)\\
\avg{\xi_{12}}^{(1)}& =& {1\over V_1 V_2}\int\int d^3\bs_1 d^3\bs_2 
{\xi^\prime(R)\over R} \bs\bx \nonumber \\
& = &
{\xi^\prime(R)\bx\over R V_1 V_2} \int\int d^3\bs_1 d^3\bs_2(\bs_2-\bs_1)
\nonumber \\
& = &
{\xi^\prime(R)\bx\over R}\left(\avg{\bs_2}-\avg{\bs_1}\right)=0.
\nonumber
\end{eqnarray}
We calculate the second order in two parts:
\begin{eqnarray}
{1\over V_1V_2}\int d\bs_1\int d\bs_2\sum s_\alpha^2 & = &
{1\over V_1V_2}\int d\bs_1\int d\bs_2 (\bs_2-\bs_1)^2
\\
& = & \avg{\bs_1^2}+\avg{\bs_2^2} = \sigma_1^2+\sigma_2^2
\nonumber \\
{1\over V_1V_2}\int d\bs_1\int d\bs_2 s_\alpha s_\beta & = &
{1\over V_1V_2}\int d\bs_1\int d\bs_2
  (s_{2\alpha}-s_{1\alpha})(s_{2\beta}-s_{1\beta})
\\
& = & \avg{s_{1\alpha}s_{1\beta}} + \avg{s_{2\alpha}s_{2\beta}}=
Q_{\alpha\beta}^{(1)}+Q_{\alpha\beta}^{(2)}
\nonumber
\end{eqnarray}
Thus, we obtain the second order term,
\begin{equation}
\avg{\xi_{12}}^{(2)}=
\left(\xi^{\prime\prime}(R) - {\xi^\prime(R)\over R}\right){1\over2R^2}
\sum\limits_{\alpha,\beta}x_\alpha x_\beta (Q_{\alpha\beta}^{(1)}+Q_{\alpha\beta}^{(2)})+
{\xi^\prime(R)\over 2R}(\sigma_1^2+\sigma_2^2),
\end{equation}
where the moments $Q_{\alpha\beta}$ and $\sigma_i$ will be evaluated
below.

Each cell has volume
\begin{equation}
V = \int d^3\bs =
\int\limits_{r_l}^{r_u}dr\int\limits_{\vartheta_l}^{\vartheta_u}d\vartheta
\int\limits_{\varphi_l}^{\varphi_u}d\varphi r^2\sin\vartheta=
(\varphi_u-\varphi_l){r_u^3-r_l^3\over3}(\cos\vartheta_l-\cos\vartheta_u).
\end{equation}
In Cartesian coordinates, the center of mass of each cell is
\begin{eqnarray}
\avg{x} & = &{1\over V}
\int\limits_{r_l}^{r_u}dr\int\limits_{\vartheta_l}^{\vartheta_u}d\vartheta
\int\limits_{\varphi_l}^{\varphi_u}d\varphi
r^2\sin\vartheta r\sin\vartheta\cos\varphi
\\
& = & {1\over V}
\left[{r^4\over4}\right]_{r_l}^{r_u}
\left[{\vartheta\over2}-{\sin2\vartheta\over 4}\right]_{\vartheta_l}
^{\vartheta_u}
\left[\sin\varphi\right]_{\varphi_l}^{\varphi_u} \nonumber \\
\avg{y} & = & {1\over V}
\int\limits_{r_l}^{r_u}dr\int\limits_{\vartheta_l}^{\vartheta_u}d\vartheta
\int\limits_{\varphi_l}^{\varphi_u}d\varphi
r^2\sin\vartheta r\sin\vartheta\sin\varphi
\\
& = & {1\over V}
\left[{r^4\over4}\right]_{r_l}^{r_u}
\left[{\vartheta\over2}-{\sin2\vartheta\over 4}\right]_{\vartheta_l}
^{\vartheta_u}
\left[-\cos\varphi\right]_{\varphi_l}^{\varphi_u} \nonumber \\
\avg{z} & = & {1\over V}
\int\limits_{r_l}^{r_u}dr\int\limits_{\vartheta_l}^{\vartheta_u}d\vartheta
\int\limits_{\varphi_l}^{\varphi_u}d\varphi
r^2\sin\vartheta r\cos\vartheta
\\
& = & {1\over V}
\left[{r^4\over4}\right]_{r_l}^{r_u}
\left[{\sin^2\vartheta\over 2}\right]_{\vartheta_l}^{\vartheta_u}
\left[\varphi\right]_{\varphi_l}^{\varphi_u}.\nonumber
\end{eqnarray}
In spherical coordinates, the center of mass is described by
\begin{eqnarray}
\tan\vartheta_c & = & {\sqrt{\avg{x}^2+\avg{y}^2}\over\avg{z}}=
{
{1\over V}\left[{r^4\over4}\right]_{r_l}^{r_u}
\left[{\vartheta\over2}-
{\sin2\vartheta\over 4}\right]_{\vartheta_l}^{\vartheta_u}
\sqrt{
\left(\left[\sin\varphi\right]_{\varphi_l}^{\varphi_u}\right)^2+
\left(\left[\cos\varphi\right]_{\varphi_l}^{\varphi_u}\right)^2}
\over
{1\over V}\left[{r^4\over4}\right]_{r_l}^{r_u}
\left[{\sin^2\vartheta\over 2}\right]_{\vartheta_l}^{\vartheta_u}
\left[\varphi\right]_{\varphi_l}^{\varphi_u}
}\\
& = &
{\vartheta_u-\vartheta_l-{1\over2}(\sin2\vartheta_u-\sin2\vartheta_l)
\over
\sin^2\vartheta_u-\sin^2\vartheta_l}\times
{\sqrt{(\sin\varphi_u-\sin\varphi_l)^2+
(\cos\varphi_u-\cos\varphi_l)^2}\over
\varphi_u-\varphi_l}\nonumber \\
& = &
\left({ \vartheta_u-\vartheta_l\over
\sin(\vartheta_u+\vartheta_l)\sin(\vartheta_u-\vartheta_l)}-
{1\over2}\cot(\vartheta_u+\vartheta_l)\right)
{\sin{\varphi_u-\varphi_l\over2}\over(\varphi_u-\varphi_l)/2}
\nonumber
\end{eqnarray}
\begin{eqnarray}
r_c^2  & = & \avg{x}^2+\avg{y}^2+\avg{z}^2
\\
& = & {1\over V^2}\left(\left[{r^4\over4}\right]_{r_l}^{r_u}\right)^2
\nonumber\\
& \times &
\left\{
\left(\left[{\vartheta\over2}-
{\sin2\vartheta\over 4}\right]_{\vartheta_l}^{\vartheta_ u}\right)^2\left[
\left(\left[\sin\varphi\right]_{\varphi_l}^{\varphi_u}\right)^2+
\left(\left[-\cos\varphi\right]_{\varphi_l}^{\varphi_u}\right)^2\right]+
\left(\left[{\sin^2\vartheta\over 2}\right]_{\vartheta_l}^{\vartheta_u}
\left[\varphi\right]_{\varphi_l}^{\varphi_u}\right)^2
\right\}
\nonumber \\
& = & {(r_u^4-r_l^4)^2\over16V^2}
\left\{
\left(\left[{\vartheta\over2}-
{\sin2\vartheta\over 4}\right]_{\vartheta_l}^
{\vartheta_u}\right)^24\sin^2{\varphi_u-\varphi_l\over2}+
\left({(\sin^2\vartheta_u-\sin^2\vartheta_l)(\varphi_u-\varphi_l)\over 2}
\right)^2
\right\}\nonumber
\end{eqnarray}
\begin{equation}
\varphi_c = {\varphi_u+\varphi_l\over2}
\end{equation}

The inertial moments $Q_{\alpha}$ and the $\sigma_i$ in equation (B13) are
\begin{eqnarray}
Q_{xx} & =  & {1\over V}\int\limits_{r_l}^{r_u}dr
\int\limits_{\vartheta_l}^{\vartheta_u}d\vartheta
\int\limits_{\varphi_l}^{\varphi_u}d\varphi r^2\sin\vartheta
r^2\sin^2\vartheta\cos^2\varphi -\avg{x}^2  \\
& = &
{1\over V}\left[{r^5\over5}\right]_{r_l}^{r_u}
\left[{\cos^3\vartheta\over3}-\cos\vartheta\right]_{\vartheta_l}^{\vartheta_u}
\left[{\varphi\over2}+{\sin2\varphi\over4}\right]_{\varphi_l}^{\varphi_u}
- \avg{x}^2
\nonumber \\
Q_{xy} = Q_{yx} & = &
{1\over V}\int\limits_{r_l}^{r_u}dr
\int\limits_{\vartheta_l}^{\vartheta_u}d\vartheta
\int\limits_{\varphi_l}^{\varphi_u}d\varphi r^2\sin\vartheta
r^2\sin^2\vartheta\sin\varphi\cos\varphi -\avg{x}\avg{y} \\
& = &
{1\over V}\left[{r^5\over5}\right]_{r_l}^{r_u}
\left[{\cos^3\vartheta\over3}-\cos\vartheta\right]_{\vartheta_l}^{\vartheta_u}
\left[{\sin^2\varphi\over2}\right]_{\varphi_l}^{\varphi_u}
-\avg{x}\avg{y}
\nonumber \\
Q_{xz} = Q_{zx} & = &
{1\over V}\int\limits_{r_l}^{r_u}dr
\int\limits_{\vartheta_l}^{\vartheta_u}d\vartheta
\int\limits_{\varphi_l}^{\varphi_u}d\varphi r^2\sin\vartheta
r^2\sin\vartheta\cos\vartheta\cos\varphi -\avg{x}\avg{z}
\\
& = &
{1\over V}\left[{r^5\over5}\right]_{r_l}^{r_u}
\left[{\sin^3\vartheta\over3}\right]_{\vartheta_l}^{\vartheta_u}
\left[\sin\varphi\right]_{\varphi_l}^{\varphi_u} -\avg{x}\avg{z}
\nonumber \\
Q_{yy} & = & {1\over V}\int\limits_{r_l}^{r_u}dr
\int\limits_{\vartheta_l}^{\vartheta_u}d\vartheta
\int\limits_{\varphi_l}^{\varphi_u}d\varphi r^2\sin\vartheta
r^2\sin^2\vartheta\sin^2\varphi -\avg{y}^2
\\
& = &
{1\over V}\left[{r^5\over5}\right]_{r_l}^{r_u}
\left[{\cos^3\vartheta\over3}-\cos\vartheta\right]_{\vartheta_l}^{\vartheta_u}
\left[{\varphi\over2}-{\sin2\varphi\over4}\right]_{\varphi_l}^{\varphi_u}
-\avg{y}^2
\nonumber \\
Q_{yz} = Q_{zy} & = &
{1\over V}\int\limits_{r_l}^{r_u}dr
\int\limits_{\vartheta_l}^{\vartheta_u}d\vartheta
\int\limits_{\varphi_l}^{\varphi_u}d\varphi r^2\sin\vartheta
r^2\sin\vartheta\cos\vartheta\sin\varphi - \avg{y}\avg{z}
 \\
& = & {1\over V}\left[{r^5\over5}\right]_{r_l}^{r_u}
\left[{\sin^3\vartheta\over3}\right]_{\vartheta_l}^{\vartheta_u}
\left[-\cos\varphi\right]_{\varphi_l}^{\varphi_u} -\avg{y}\avg{z}
\nonumber \\
Q_{zz} & = & 
{1\over V}\int\limits_{r_l}^{r_u}dr
\int\limits_{\vartheta_l}^{\vartheta_u}d\vartheta
\int\limits_{\varphi_l}^{\varphi_u}d\varphi r^2\sin\vartheta
r^2\cos^2\vartheta - \avg{z}^2 \\
& = &
{1\over V}\left[{r^5\over5}\right]_{r_l}^{r_u}
\left[-{\cos^3\vartheta\over3}\right]_{\vartheta_l}^{\vartheta_u}
\left[\varphi\right]_{\varphi_l}^{\varphi_u} -\avg{z}^2
\nonumber
\end{eqnarray}
\begin{equation}
\sigma^2=\avg{\bs^2} = \avg{x^2+y^2+z^2} -\avg{x}^2-\avg{y}^2-\avg{z}^2=
Q_{xx}+Q_{yy}+Q_{zz}
\end{equation}

%% end of appendices

\begin{figure}
\figurenum{1}
\plotone{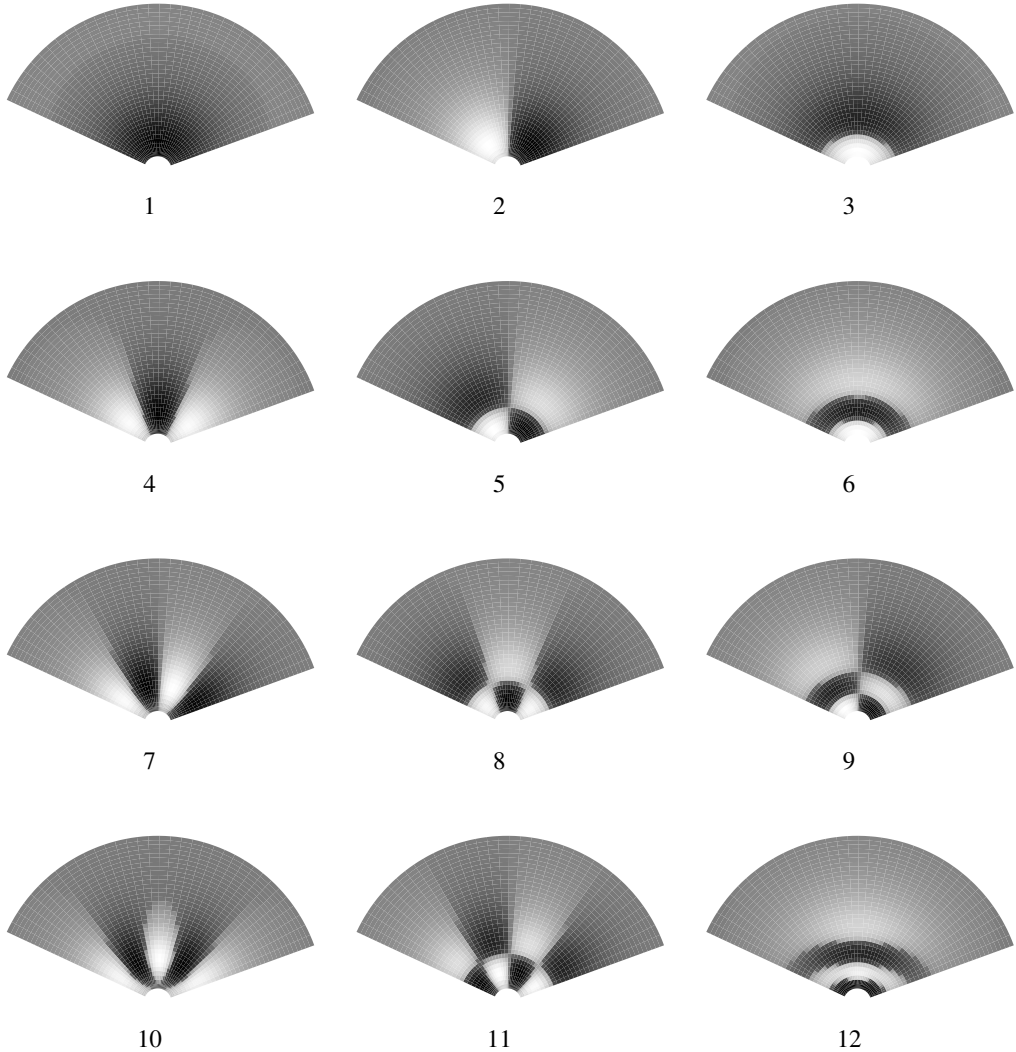}
\caption{
Eigenmodes of the apparent-magnitude limited CfA slice,
formed by assuming the selection function
and power spectrum measured for the CfA2 survey. 
This slice covers the region
$29\fdg 5\leq\delta\leq 32\fdg 5$, $8^h\leq \alpha\leq 17^h$,
and we restrict the redshift range to $10h^{-1}{\mathrm{Mpc}}\leq r\leq
120h^{-1}{\mathrm{Mpc}}$.
We plot the twelve modes with largest expected signal-to-noise ratio.
These functions closely resemble the multipole moments of the density
field, and are most sensitive to structure near the peak of the
redshift distribution $r\sim 55h^{-1}$Mpc.
}
\end{figure}

\begin{figure}
\figurenum{2}
\plotone{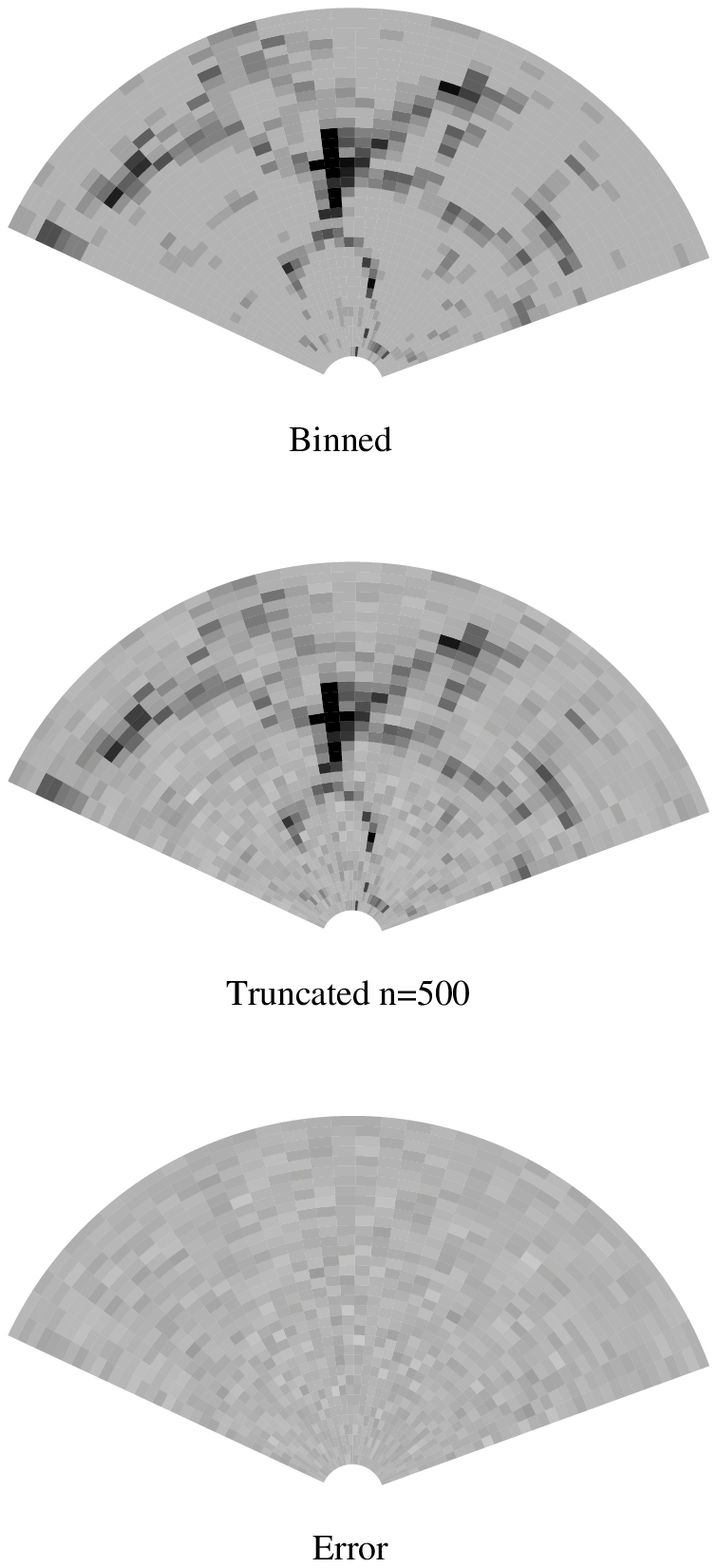}
\caption{
Demonstration of the optimal representation property of 
the K-L transform. From top to bottom, these figures show
(a) the binned
distribution of galaxies in the CfA slice (de Lapparent, Geller, \& Huchra
1986), 
(b) these same data, represented by the first 500 eigenmodes (truncation
at a signal-to-noise ratio of unity -- see Fig. [3]), and
(c) the error caused by this truncation, which is the difference between
images (a) and (b).
}
\end{figure}

\begin{figure}
\figurenum{3}
\plotone{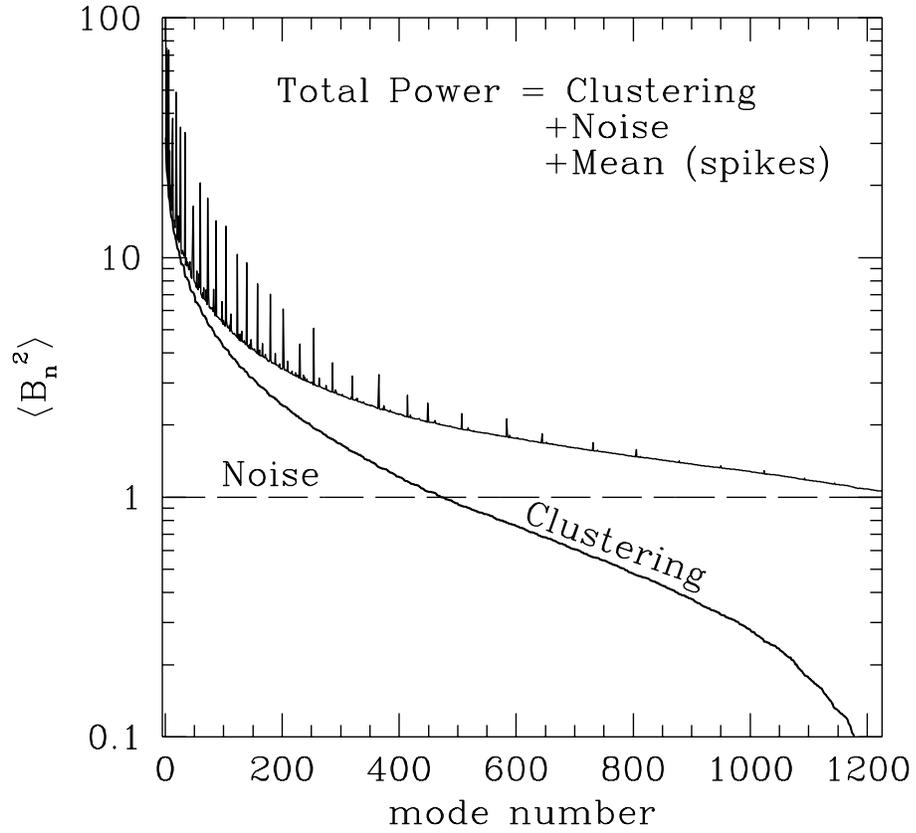}
\caption{
Expectation value of the power per mode for the K-L expansion
of the CfA slice, where the modes are ordered by decreasing signal-to-noise
ratio.
The total power (upper solid line)
is the sum of the clustering signal (lower
solid line), noise (long-dashed line), and the mean density (spikes in the
total power curve).
The expected signal-to-noise ratio is less than unity
for $n\simgreat 500$.
}
\end{figure}

\begin{figure}
\figurenum{4}
\plotone{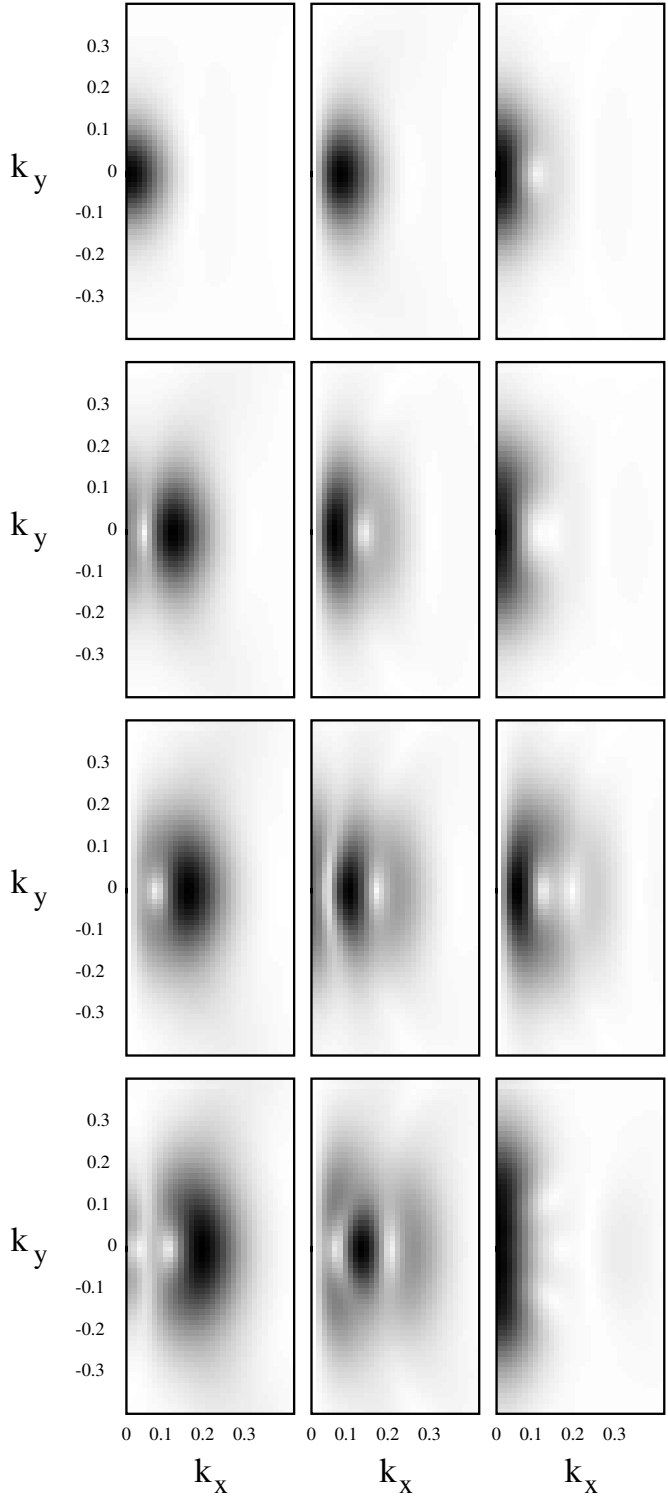}
\caption{
2-D Fourier window functions $|G_n(\kbf)|^2$ (see eq.[32] 
in section 3.4) of the twelve eigenmodes shown in
Fig. 1, computed by restricting the Fourier transform to modes in a plane
tangent to the CfA slice. 
Note that only certain of the modes are sensitive to the mean density.
For example, $n=1$ samples power near $k=0$, but $n=2$ has no sensitivity
to the mean density.
Successive modes sample power at generally larger wavenumber.
}
\end{figure}

\begin{figure}
\figurenum{5}
\plotone{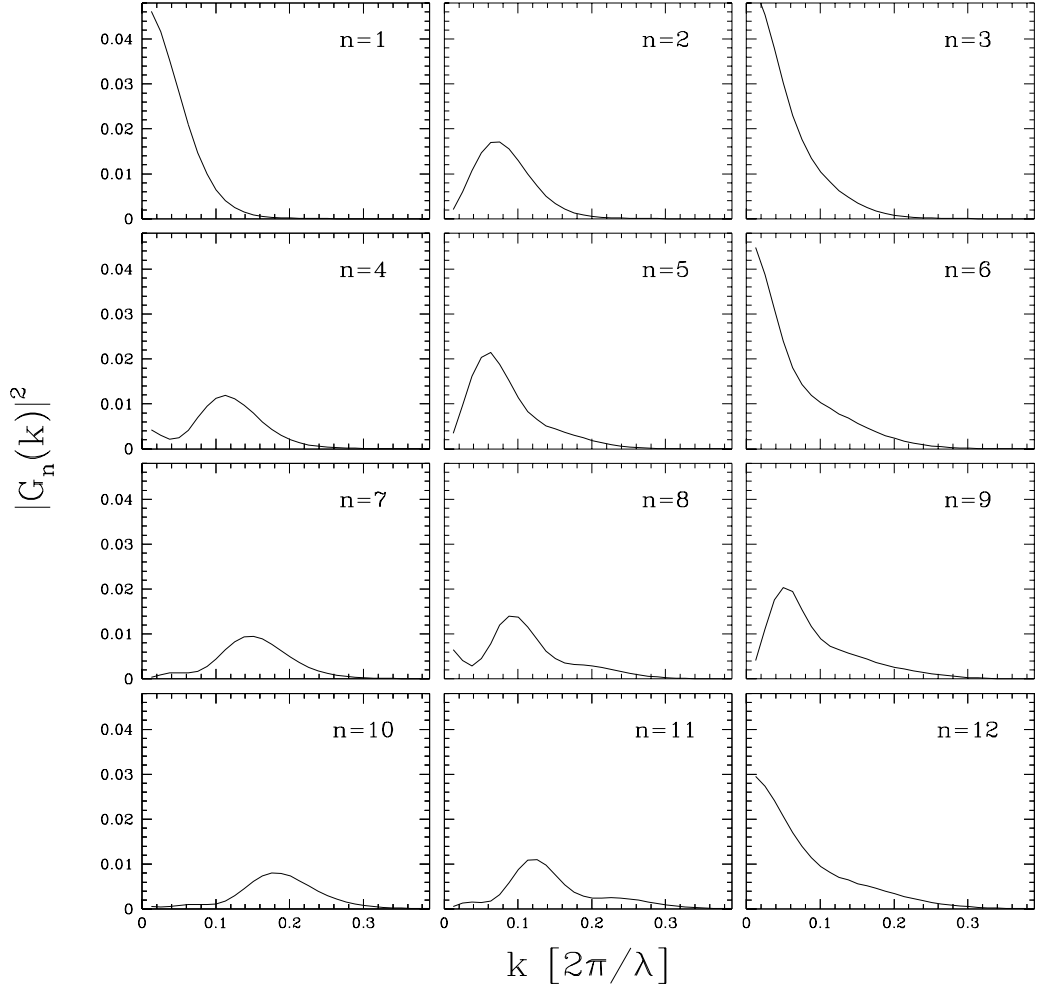}
\caption{
Fourier windows of the eigenmodes, 
as in Fig. 4, but averaged over all angles
$\hat{\kbf}$ to indicate the band of $k-$space sampled by each mode.
}
\end{figure}

\begin{figure}
\figurenum{6}
\plotone{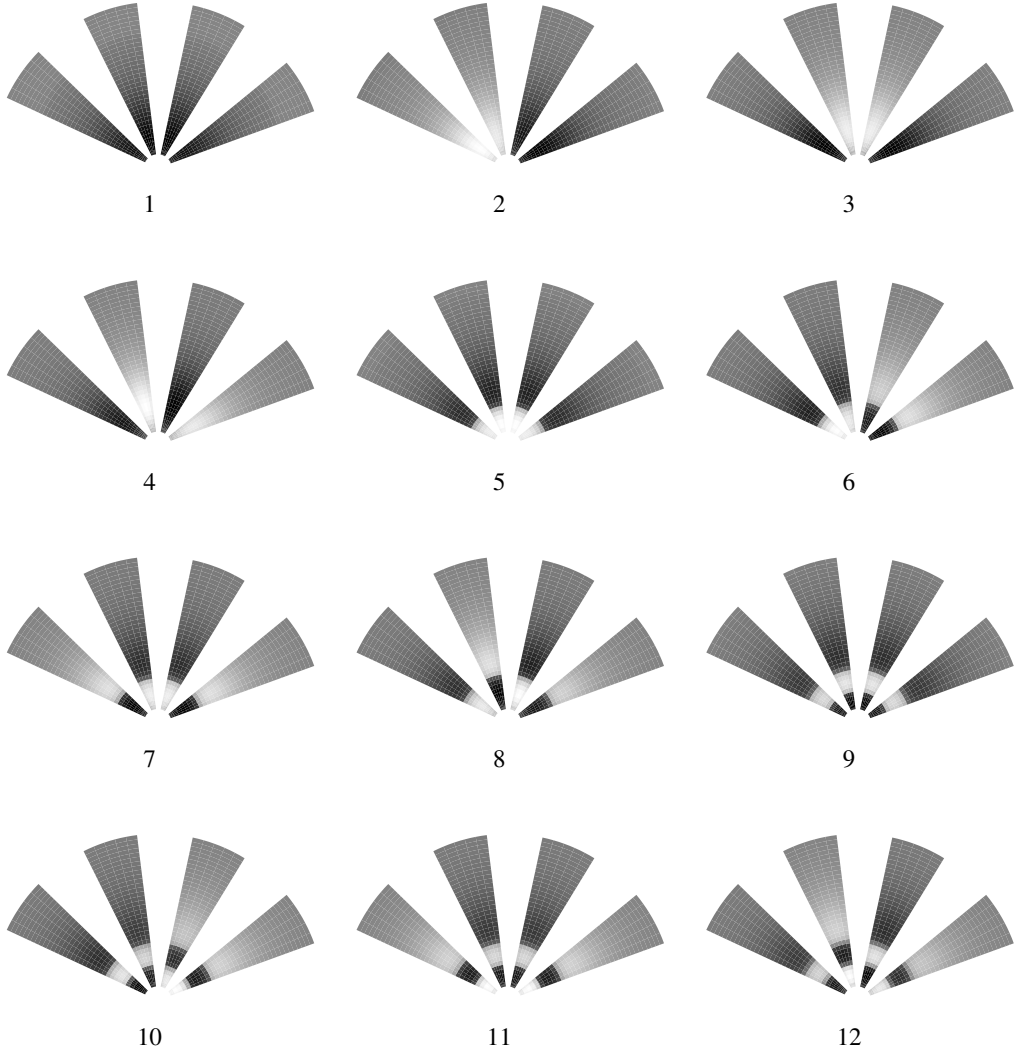}
\caption{
Eigenmodes of a survey comprised of four
narrow beams within the CfA slice.
We plot the twelve modes with largest expected signal-to-noise ratio.
Compare with Fig. 1 and note the similarity in modes that sample
large-scale density fluctuations (compare modes 1-6 in this figure
with modes 1, 2, 4, 7, 3, and 5, respectively, in Fig. 1).
}
\end{figure}

\end{document}